\documentclass[twocolumn,twoside,slac_two]{revtex4}
\usepackage{graphicx}
\usepackage{fancyhdr}
\def\Dzbpar  {\ensuremath{\Dbar^{(*)0}}\xspace}
\def\CP                {\ensuremath{C\!P}\xspace}
\def\CPp                {\ensuremath{C\!P\!+}\xspace}
\def\CPm                {\ensuremath{C\!P\!-}\xspace}

\newcommand{\bdkm}{$B^{-}\to DK^{-}$}
\newcommand{\bdkp}{$B^{+}\to DK^{+}$}

\newcommand{\bdskm}{$B^{-}\to D^{*}K^{-}$}
\newcommand{\bdskp}{$B^{+}\to D^{*}K^{+}$}

\newcommand{\bdksm}{$B^{-}\to DK^{*-}$}
\newcommand{\bdksp}{$B^{+}\to DK^{*+}$}
\def \rb {\ensuremath {r_B}\xspace}

\def \rbs {\ensuremath {r^\ast_B}\xspace}

\def \rs {\ensuremath {r_s}\xspace}

\def\dodstartilde {\ensuremath {\tilde{D}^{(\ast)0}}\xspace}
\def\Dztilde   {\ensuremath {\tilde{D}^0}\xspace}
\def\dstpi{{\Dstar\pi}}

\def \btodsospi {\mbox{$B^0\to D_{s}^{(*)+}\pi^-$}}
\def \btodospi{\ensuremath{B^0\to D^{(*)-}\pi^+}}

\def\DDstarz {\ensuremath{D^{(*)0}}\xspace}
\def\DDstarzb{\ensuremath{\Dbar^{(*)0}}\xspace}
\def\stwobg   {\ensuremath{\sin(2\beta+\gamma)}\xspace}
\def\rbk {\ensuremath{\tilde{r}_B}}
\def\brscale{\ensuremath{\times10^{-5}}}
\def\Btilde {\ensuremath{\tilde{B}}}
\def\Bztilde {\ensuremath{\Btilde^{0}}}
\def\Ktilde {\ensuremath{\tilde{K}}}

\def\Kztilde  {\ensuremath{\Ktilde^{0}}\xspace}

\input babarsym

\begin{document}

\title{Measurement of the CKM angles $\gamma$ and $2\beta+\gamma$ at B factories}

%

\author{Jean-Pierre LEES}
\affiliation{Laboratoire d'Annecy-le-Vieux de Physique des
Particules (LAPP), IN2P3/CNRS et Universit\'e de Savoie, 9 Chemin
de Bellevue, BP 110, F-74941 Annecy-le-Vieux CEDEX - FRANCE }

\begin{abstract}
On behalf of the \babar\ and Belle collaborations, we report on
the measurement of the  angle $\gamma$ and on the sum of angles
$2\beta+\gamma$ of the Unitarity Triangle.
\end{abstract}

\maketitle

\thispagestyle{fancy}


\section{Introduction}
The angle $\gamma$ (or $\phi_3$) of the unitarity triangle is
related to the complex phase of the CKM matrix element $V_{ub}$
through $V_{ub}=|V_{ub}|e^{-i\gamma}$. Various methods have been
proposed to measure the angle $\gamma$. We report on two classes
of measurements: time independent measurements in decays $B^\pm
\to \Dz/\Dzb K^{\pm}$ exploit the interference between the $b\to c
\bar u s$ and the $b\to u \bar c s$ decay amplitudes and are
sensitive to the angle $\gamma$; time dependent asymmetries in
decays $B^0 \to D^{(*)\pm} \pi^{\mp}$ or $B^0 \to D^{0} K^{0}$
occur through the interference of the favored amplitude $\Bz \to
\Dbar^{(*)-}\pi^+$ ($\Bz \to \Dbar^{(*)0}K^0$) and the suppressed
amplitude $\Bzb \to \Dbar^{(*)-}\pi^+$ ($\Bzb \to
\Dbar^{(*)0}K^0$) plus \Bz \Bzb mixing, and are sensitive to the
sum of angles $2\beta+\gamma$.

\section{Measurement of $\gamma$ in $B^\pm \to \Dz/\Dzb K^{\pm}$}

\begin{figure}[hb]
\centering
\includegraphics[width=40mm]{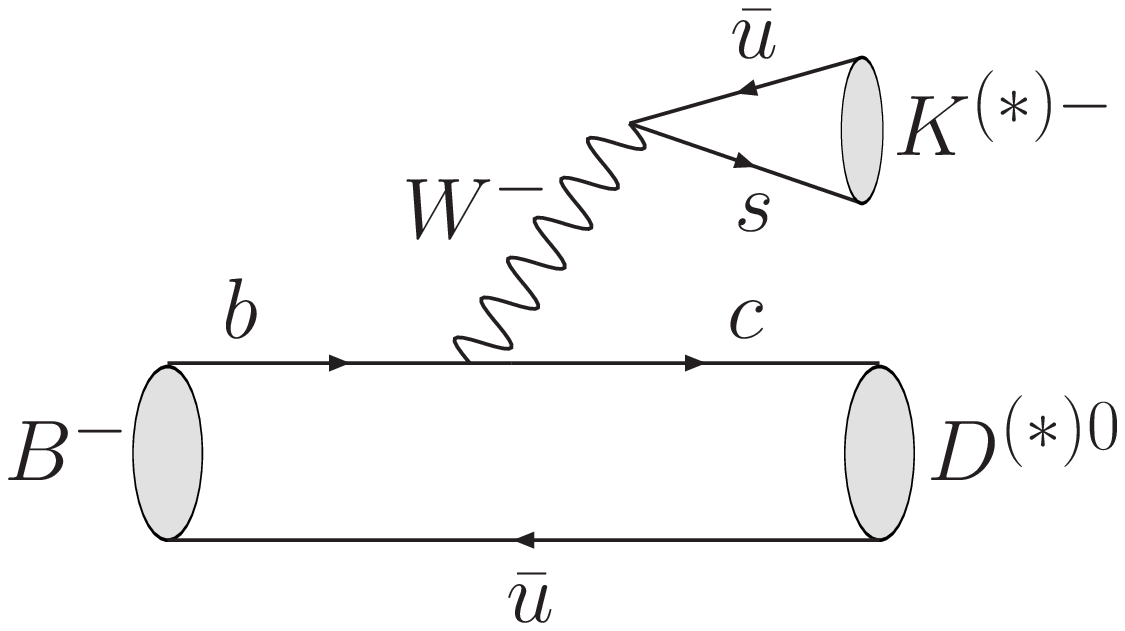}
\includegraphics[width=40mm]{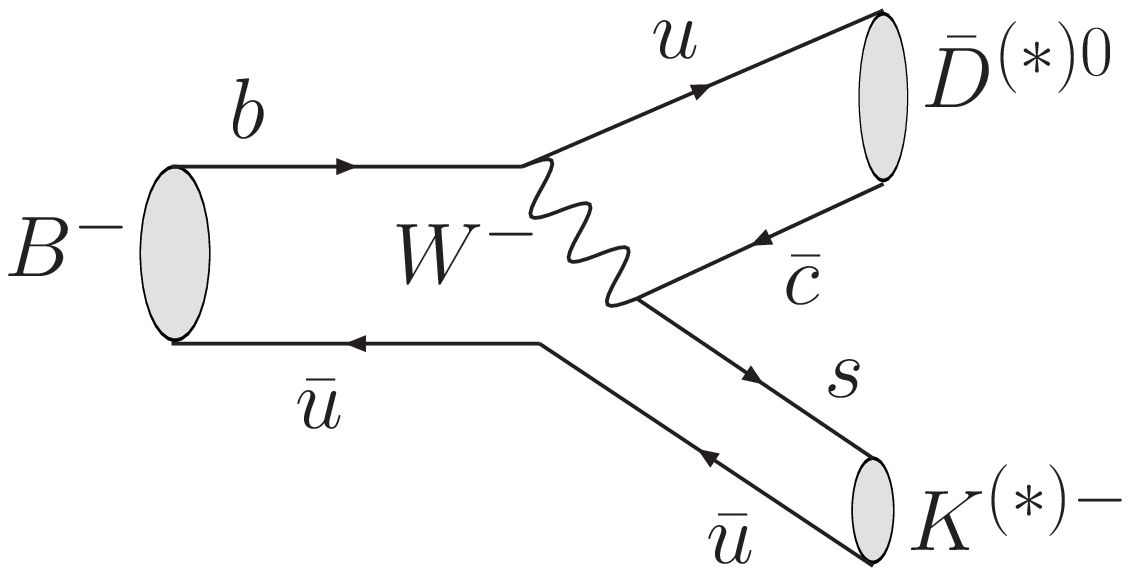}
  \caption{Feynman diagrams for $B^- \to D^{(*)0} K^{(*)-}$ and $\Dzbpar K^{(*)-}$.
   The latter is CKM and color suppressed with respect to the former.
   The CKM-suppression factor is
   $|V_{ub}V_{cs}^*/V_{cb}V_{us}^*| \approx 0.4$.  The naive
   color-suppression factor is $\frac{1}{3}$.}
\label{fig:feynmandk}
 \end{figure}

The experimental techniques used to measure $\gamma$ in charged
$B$ decays exploit the interference between $B^- \to
D^{(*)0}K^{(*)-}$ and $B^- \to \Dzbpar K^{(*)-}$
(Fig.~\ref{fig:feynmandk}) that occurs when the $D^{(*)0}$ and the
   \Dzbpar decay to common final states.

Three different methods have been used so far:

\begin{itemize}
\item The GLW method \cite{ref:GLW}: the \Dz and the \Dzb decay to
a CP eigenstate \item The ADS method \cite{ref:ADS}: the \Dz from
the favored $\b\to c$ amplitude is reconstructed in the
doubly-Cabbibo suppressed final state $K^+\pi^-$ , while the \Dzb
from the $\b\to u$ suppressed amplitude is reconstructed in the
favored final state $K^+\pi^-$.
 \item The GGSZ (Dalitz) method \cite{ref:GGSZ}: the \Dz and the
 \Dzb are reconstructed in the same $K^0_S \pi^+\pi^-$ three body final
 state. This is based on the analysis of the $K^0_S \pi^+\pi^-$ Dalitz
 distribution and can to some extent be considered as a mixture of
 the two previous methods, depending on the position in the Dalitz
 plot.
\end{itemize}

In all three methods, the experimental observables depend on two
additional parameters which need to be determined in order to
extract useful constraints on the value of $\gamma$: the magnitude
\rb of the ratio of the amplitudes for the processes $\Bm\to \Dzb
K^-$ and $\Bm \to \Dz K^-$ (Fig.~\ref{fig:feynmandk}) and the
relative strong phase $\delta_B$ between these two amplitudes. The
amplitude ratio \rb and the phase $\delta_B$ are specific of each
$B$ decay mode reconstructed ($D^0K$, $D^{*0}K$ and $D^0K^{*}$).

\subsection{The GLW method and results:}

\begin{table*}[tb]
  \centering
  \caption{Summary of \babar\  and Belle measurements of the GLW observables $R_{CP}$ and $A_{CP}$.}\label{tab:glw}
\begin{tabular}{|c|c|c|c|c|c|c|}
  \hline
  Mode & Experiment & $N(B\overline B)$ [$10^6$]& $A_{CP+}$ & $A_{CP-}$ & $R_{CP+}$ & $R_{CP-}$ \\
  \hline
  $B\to D^0K$ & \babar \cite{pap:babardkglw}      & 232 & $+0.35 \pm 0.13 \pm 0.04$  & $-0.06 \pm 0.13 \pm 0.04$  & $0.90 \pm 0.12 \pm 0.04$  & $0.86 \pm 0.10 \pm 0.05$ \\
              & Belle  \cite{pap:belledkglw}      & 275 & $+0.06 \pm 0.14 \pm 0.05$  & $-0.12 \pm 0.14 \pm 0.05$  & $1.13 \pm 0.16 \pm 0.08$  & $1.17 \pm 0.14 \pm 0.14$  \\
              & HFAG Average \cite{web:HFAG}  &  -  & $+0.22 \pm 0.10$  & $-0.09 \pm 0.10$  & $0.90 \pm 0.10$  & $0.94 \pm 0.10$  \\
  \hline
 $B\to D^{*0}K$  & \babar \cite{pap:babardstkglw} & 123 & $-0.10\pm 0.23 ^{+0.03}_{-0.04} $  & -  & $1.06\pm0.26^{+0.10}_{-0.09}$  & -  \\
                 & Belle \cite{pap:belledkglw}    & 275 & $-0.20 \pm 0.22 \pm 0.04$ & $+0.13 \pm 0.30 \pm 0.08$  & $1.41 \pm 0.25 \pm 0.06$  & $1.15 \pm 0.31 \pm 0.12$  \\
                 & HFAG Average \cite{web:HFAG} &  -  & $-0.15 \pm 0.16$          & $+0.13 \pm 0.31$  & $1.25\pm 0.19$  & $1.15 \pm 0.33$  \\
  \hline
  $B\to D^0K^*$ & \babar \cite{pap:babardksglw}        & 232 & $-0.08 \pm 0.19 \pm 0.08$  & $-0.26 \pm 0.40 \pm 0.12$  & $1.96 \pm 0.40 \pm 0.11$  & $0.65 \pm 0.26 \pm 0.08$  \\
                & Belle         & -  & -  & -  & -  &   \\
                & HFAG Average \cite{web:HFAG} & -  &  $-0.08 \pm 0.21$ &   $-0.26 \pm 0.42$  &  $1.96 \pm 0.41$  &   $0.65 \pm 0.27$   \\
  \hline
\end{tabular}
\label{tab:glw}
\end{table*}

\begin{figure}[h]
  \begin{center}
    \includegraphics[width=4.0cm]{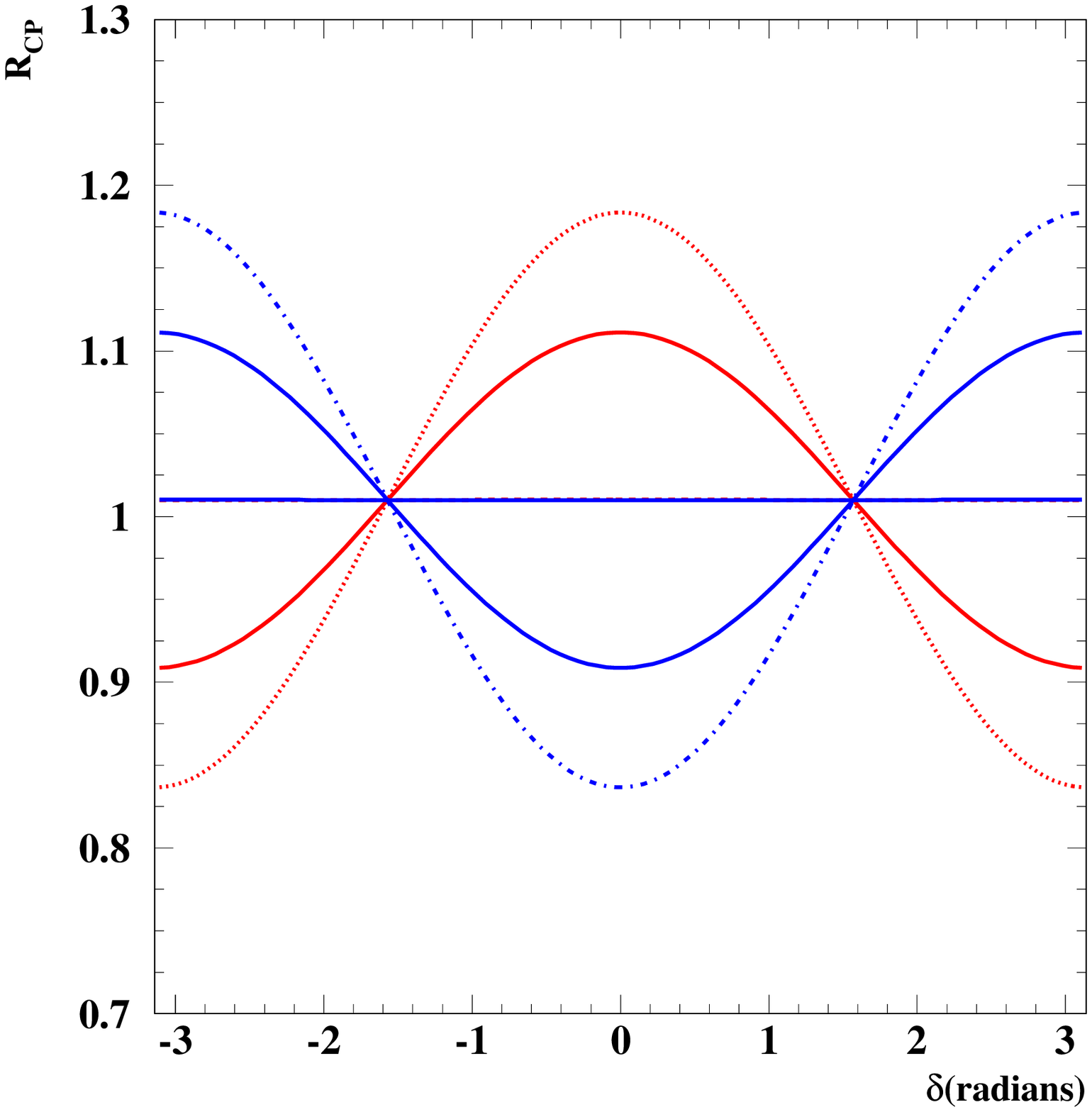}
    \includegraphics[width=4.0cm]{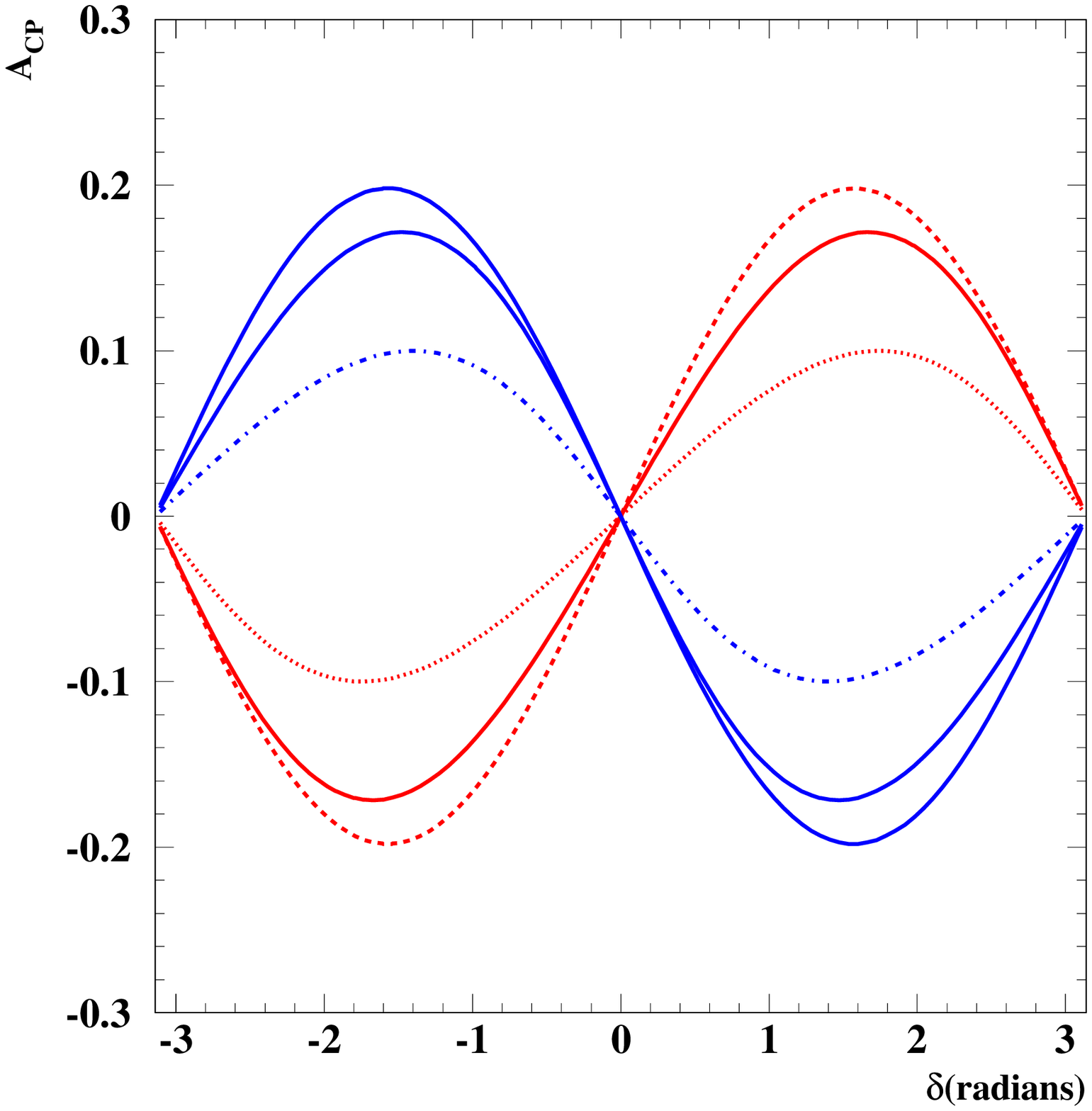}
    \caption{ The variation of $R_{CP}$ (left) and $A_{CP}$ (right) vs the strong phase $\delta_B$ for $r_B=0.10$ and
    $\gamma = 30^o$, $60^o$ and $90^o$. Red is for \CPp modes and blue for \CPm modes.}
    \label{fig:asym01}
  \end{center}
\end{figure}

\begin{figure}[!htb]
\begin{center}
\includegraphics[width=7.5cm,height=3.6cm]{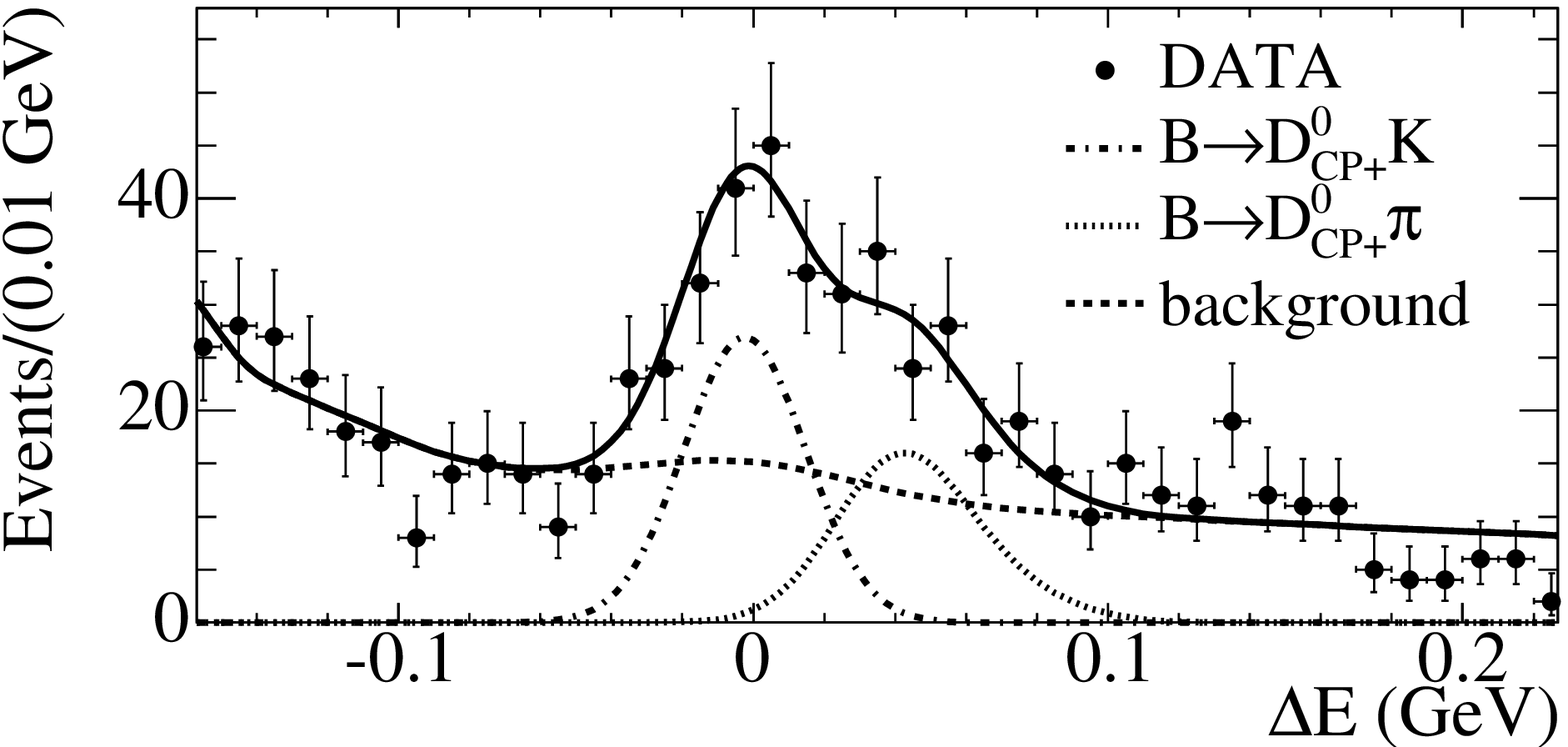}
\includegraphics[width=7.5cm,height=3.6cm]{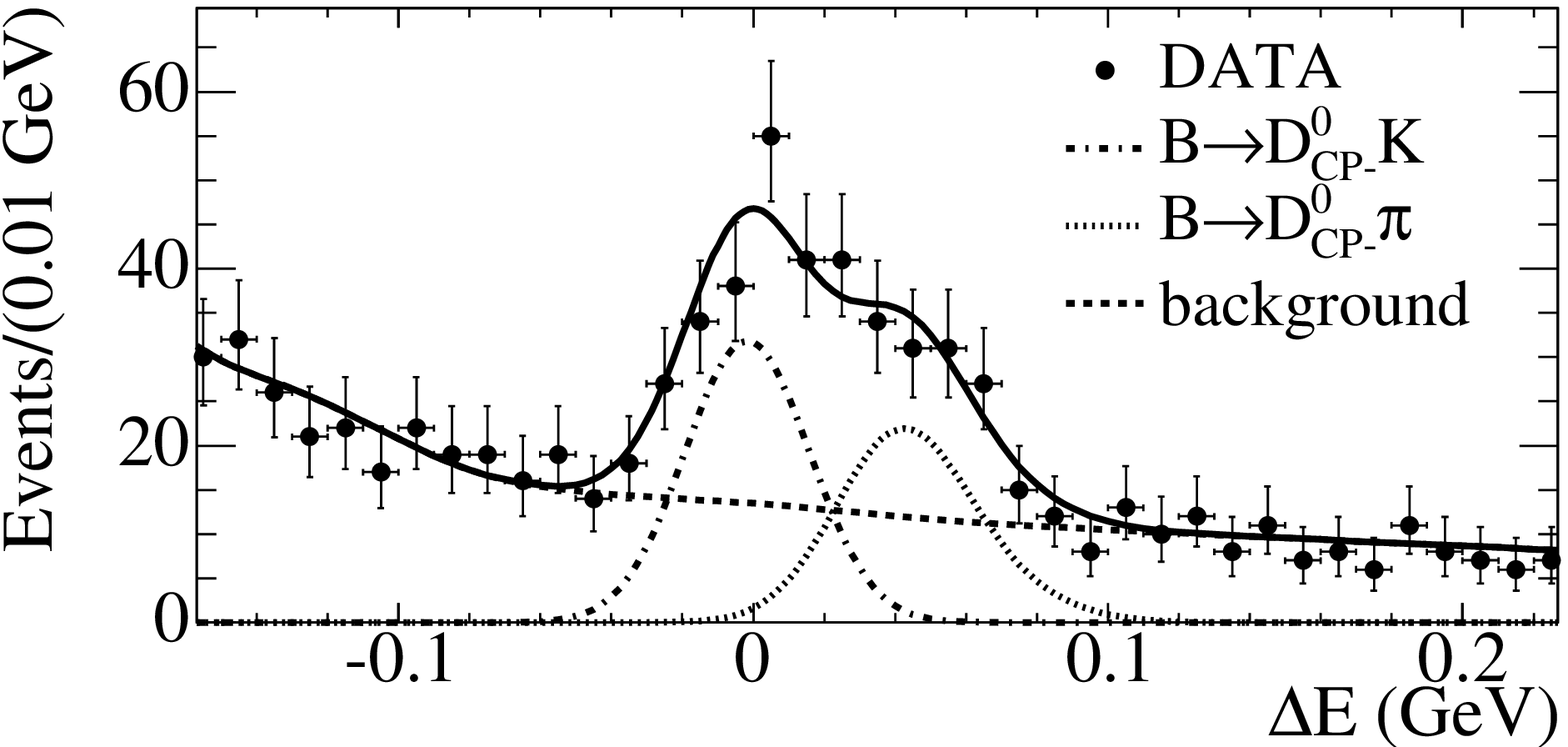}
\caption{Distributions of \DeltaE\ for events enhanced in
  $B{\to}D^0K$ signal at \babar. Top:  $B^-{\to}D^0_{\CPp}K^-$; bottom: $B^-{\to}D^0_{\CPm}K^-$. Solid
curves
  represent projections of the maximum likelihood fit; dashed,
  dashed-dotted and dotted curves represent the $B{\to}D^0K$,
  $B{\to}D^0\pi$ and background contributions.}
\label{fig:fit_kaons}
\end{center}
\end{figure}

The results of the GLW analyses are usually expressed in terms of
the ratios $R_{\CP\pm}$ of charge-averaged partial rates and of
the partial-rate charge asymmetries $A_{\CP\pm}$,
\begin{eqnarray}
  &&R_{\CP\pm} = \frac{\Gamma(B^-{\to}\Dz_{\CP\pm}K^-) +
    \Gamma(B^+{\to}\Dz_{\CP\pm}K^+)} {\left[\Gamma(B^-{\to}\Dz K^-)+\Gamma(B^+{\to}\Dzb K^+)\right]/2}\,,\ \ \ \  \\
&&A_{\CP\pm}=\frac{\Gamma(B^-{\to}\Dz_{\CP\pm}K^-)-\Gamma(B^+{\to}\Dz_{\CP\pm}K^+)}{\Gamma(B^-{\to}\Dz_{\CP\pm}K^-)+\Gamma(B^+{\to}\Dz_{\CP\pm}K^+)}\,.\
\ \ \
\end{eqnarray}
where \CPp refers to the CP-even final states $\pi^+\pi^-$ and
$K^+K^-$ and \CPm refers to the CP-odd final states $\KS\piz$,
$\KS\phi$ and $\KS\omega$. $R_{\CP\pm}$ and $A_{\CP\pm}$ are
related to the angle $\gamma$, the amplitude ratio $r_B$ and the
strong phase difference $\delta_B$ through the relations
$R_{\CP\pm}=1+r_B^2\pm 2r_B\cos\delta_B\cos\gamma$ and
$A_{\CP\pm}=\pm
2r_B\sin\delta_B\sin\gamma/R_{\CP\pm}$~\cite{ref:GLW}, thus
allowing a determination of the 3 unknowns ($r_B$, $\delta_B$ and
$\gamma$) up to an 8 fold ambiguity in $\gamma$. The variation of
$R_{CP}$ and $A_{CP}$ vs the strong phase $\delta_B$ for
$r_B=0.10$ is shown in Fig.\ref{fig:asym01} for different values
of $\gamma$. The asymmetries for \CPp and \CPm states have
opposite signs, while the Bf ratios are approximately symmetric
respective to 1. The GLW method is theoretically clean, with
nearly no hadronic uncertainty. However, it is experimentally
challenging, as the effective branching ratio for the decay modes
reconstructed is of the order of $10^{-6}$.
 Both \babar\ and Belle reconstruct the CP-even and CP-odd modes listed above and have published recently final
results using statistical samples larger than 200 million
$B\overline B$ events. The corresponding numbers of reconstructed
events are about 100 \CPp and 100 \CPm events in each experiment
for the $B^-\to D^0K^-$ modes, and even less for the modes $B^-\to
D^{*0}K^-$ and $B^-\to D^0K^{*-}$, which have a lower
reconstruction efficiency. The $\Delta E$ distribution of \CPp and
\CPm events in the \babar\ $B^-\to D^0K^-$ analysis is shown in
Fig.\ref{fig:fit_kaons}. The different \babar\ and Belle $B^-\to
D^{(*)0}K^{(*)-}$ analysis are described in detail in
references~\cite{pap:babardkglw, pap:babardksglw,
pap:babardstkglw,pap:belledkglw} and their results are summarized
in Table~\ref{tab:glw}. Due to the limited statistics and to the
smallness of the \rb parameter, the GLW method alone is not yet
able to provide strong constraints on $\gamma$. For the
\Bm\to\Dz\Km decay channel, 3$\sigma$ significant differences
between \Bm and \Bp data seem to be within reach in the near
future, when $\sim 1 ab^{-1}$ of data will have been collected in
each experiment.

\subsection{ADS Results:}
\begin{figure}[h]
  \begin{center}
    \includegraphics[width=7.5cm]{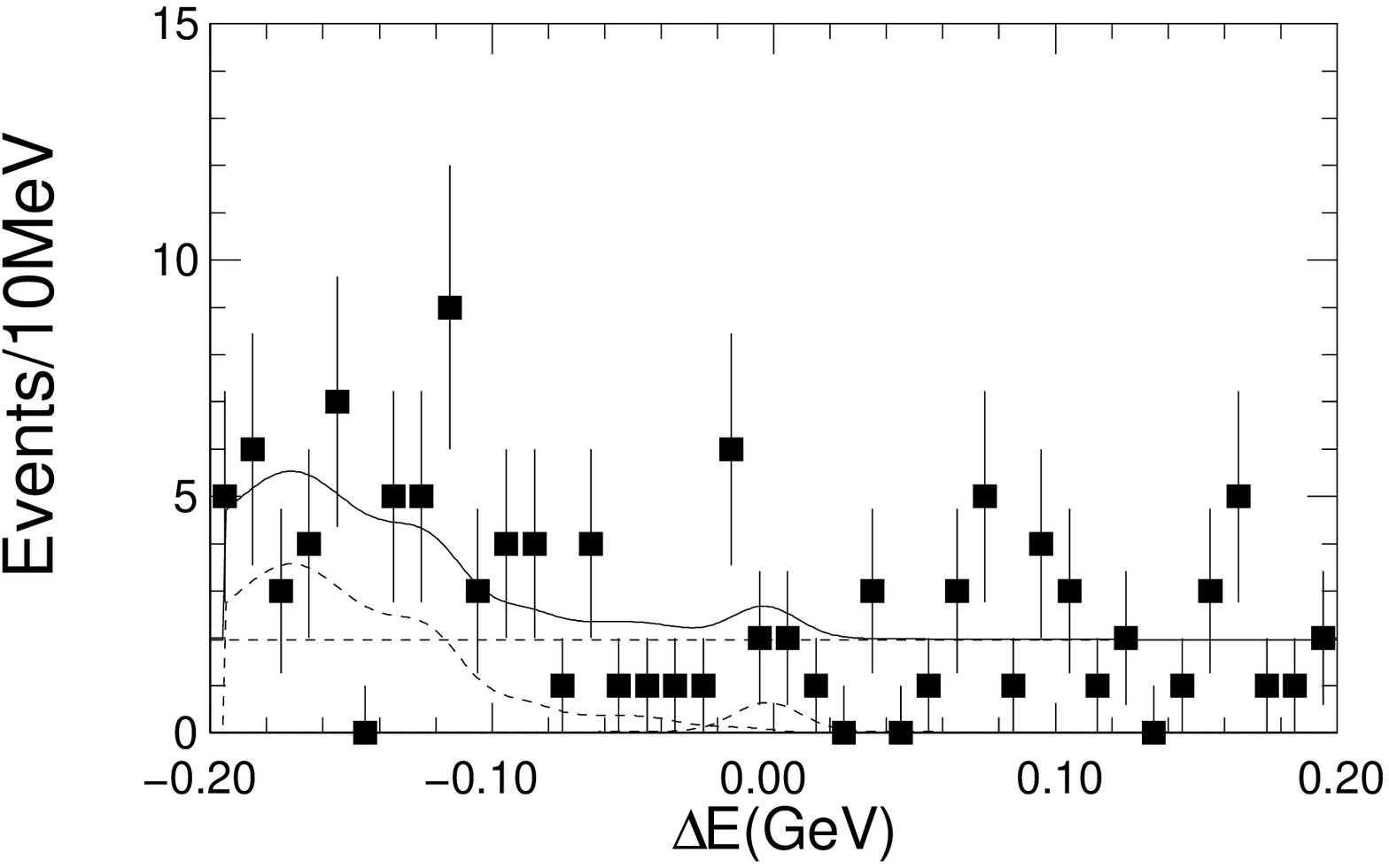}
    \includegraphics[width=7.5cm]{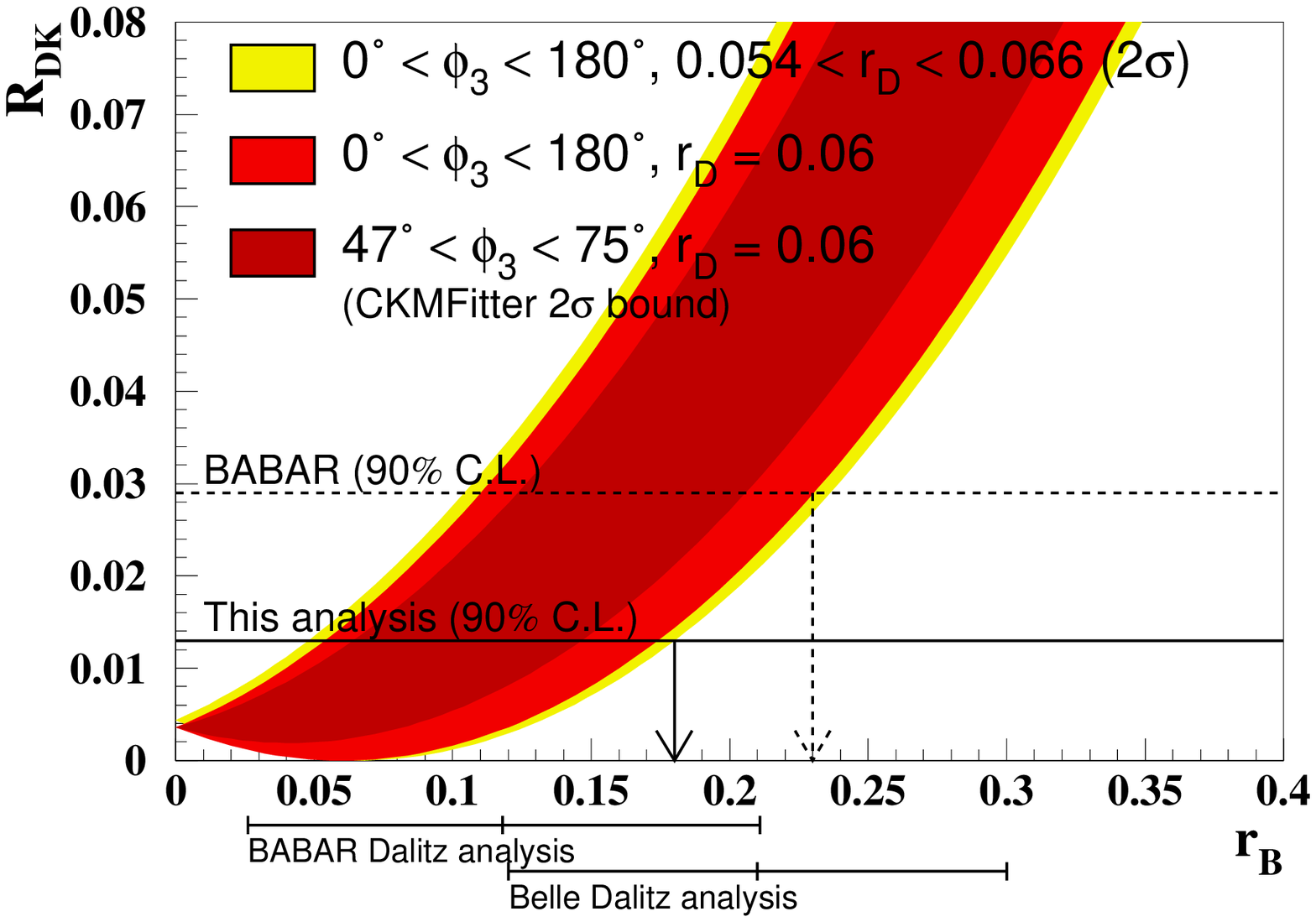}
    \caption{(Top) Belle $\Delta E$ fit results for $B^- \to D_{sup}K^-$. (Bottom) Belle constraint on $r_B$ from $R_{DK}$.}
    \label{fig:rb_rdk}
  \end{center}
\end{figure}

\begin{table*}[tb]
\begin{center}
\caption{Summary of \babar\ and Belle ADS measurements.}
\begin{tabular}{|c|c|c|c|c|c|c|}
 \hline
   Mode & Experiment & $N(B\overline B)$ [$10^6$]& $R_{ADS}$ & $R_{ADS}$ & $r_B$ & $r_B$  \\
       &            &                           &            & 90\% C.L.limit&  & 90\% C.L.limit \\ \hline
  $B\to D^0K$ & \babar \cite{pap:babarads1}                 & 232 & $0.013^{+0.011}_{-0.009}$  & $<0.029 $  && $r_B<0.23$  \\
              & Belle  \cite{pap:belleads}                  & 386 & $0.000\pm 0.008\pm 0.001$  & $<0.0139$  && $r_B<0.18$  \\ \hline
 $B\to D^{*0}_{(D^0\pi^0)}K$  &\babar \cite{pap:babarads1}  & 232 & $-0.002^{+0.010}_{-0.006}$ & $<0.023 $  && $r_B^{*2}<(0.16)^2$ \\
 $B\to D^{*0}_{(D^0\gamma)}K$ &                             &     & $0.011^{+0.018}_{-0.013}$  & $<0.045 $  &&   \\
  \hline
  $B\to D^0K^*$ & \babar \cite{pap:babarads2}             & 232 & $0.046\pm 0.031\pm 0.08$   & $       $  & $r_B=0.28^{+0.006}_{-0.010}$ &\\
  \hline

\end{tabular}
\label{tab:ads}
\end{center}
\end{table*}
In the Atwood-Dunietz-Soni (ADS) method, instead of using \CP
eigenstate decays of the \Dz, the decays \Dz \to \Kp \pim and \Dzb
\to \Kp \pim are used. The overall effective branching ratio for
the final state $\Bm \to [\Kp \pim]_{D^0} \Km$ is expected to be
small ($\sim 10^{-7}$), but the two interfering diagrams are of
the same order of magnitude and large asymmetries are therefore
expected. The favored decay mode $\Bm \to [\Km \pip]_{D^0} \Km$ is
used to normalize the measurement and cancel many experimental
systematics. The main experimental observable are the ratio
$R_{ADS}$ of the suppressed to favored modes and the \Bm / \Bp
asymmetry:
\begin{eqnarray}
  R_{ADS} & = & \frac{ {\cal B}([\Kp \pim]_{D^0}\Km) + {\cal B}([\Km \pip]_{D^0}\Kp) }
                     { {\cal B}([\Km \pip]_{D^0}\Km) + {\cal B}([\Kp \pim]_{D^0}\Kp)
                     } \nonumber \\
                     & = & r_D^2 + 2 r_D r_B \cos \gamma
                     \cos(\delta_B+\delta_D)+r_B^2 \label{eq-rads}
                     \\
  A_{ADS} & = & \frac{ {\cal B}([\Kp \pim]_{D^0}\Km) - {\cal B}([\Km \pip]_{D^0}\Kp) }
                     { {\cal B}([\Kp \pim]_{D^0}\Km) + {\cal B}([\Km \pip]_{D^0}\Kp)
                     } \nonumber \\
                     & = & 2 r_D r_B \sin \gamma
                     \sin(\delta_B+\delta_D)/R_{ADS}
                     \label{eq-aads},
                     \end{eqnarray}
where $r_B = |A(\Bm \to \Dzb \Km) / A(\Bm \to \Dz \Km)|$ and $r_D
= |A(\Dz \to \Kp \pim) / A(\Dz \to \Km \pip)| = 0.060\pm
0.003$\cite{pap:babarrd} are the suppressed to favored $B$ and $D$
amplitude ratios, and $\delta_B$ and $\delta_D$ are the strong
phase differences between the two $B$ and $D$ decay amplitudes,
respectively. As it can be seen from Eq. \ref{eq-rads}, $R_{ADS}$
is highly sensitive to $r_B^2$. Using a sample of $232 \times
10^6$ \BB events, \babar\ reconstructs $5^{+4}_{-3}$ events in the
\Bm\to \Dz[\Kp\pim] \Km channel, $-0.2^{+1.3}_{-0.7}$ events in
the \Bm\to \Dstarz[\Kp\pim] \Km channel (\Dstarz\to\Dz\piz),
$1.2^{+2.1}_{-1.4}$ events in the \Bm\to \Dstarz[\Kp\pim] \Km
channel (\Dstarz\to\Dz\g), and $4.2\pm 2.8$ events in the \Bm\to
\Dz[\Kp\pim] \Kstarm channel \cite{pap:babarads1, pap:babarads2}.
From $386 \times 10^6$ \BB events, Belle reconstructs
$0.0^{+5.3}_{-5.0}$ events in the \Bm\to \Dz[\Kp\pim] \Km channel
\cite{pap:belleads}. None of these results are statistically
significant and for the $\Dz K$ and $\Dstarz K$ channels limits on
$R_{ADS}$ and $r_B$ are extracted. The Belle result for \Bm \to
\Dz \Km is illustrated in Fig.\ref{fig:rb_rdk}. The least
restrictive limit is obtained allowing $\pm 1\sigma$ variation on
$r_D$ and assuming maximal interference ($\gamma=0^o$,
$\delta_B+\delta_D=180^o$ or $\gamma=180^o$,
$\delta_B+\delta_D=0^o$) and is found to be $r_B<0.18$ at
90\%C.L..   For the \babar\ \Bm\to \Dz[\Kp\pim]\Kstarm result, a
frequentist approach has been used to combine the results from the
GLW and ADS methods, resulting in $r_B=0.28^{+0.06}_{-0.10}$ and
excluding the interval $75^o<\gamma<105^o$ at the two standard
deviation level. A summary of the \babar\ and Belle ADS results
can be found in Table~\ref{tab:ads}, and more details on the
analysis in Ref.\cite{pap:babarads1, pap:babarads2, pap:belleads}.
Similar to the GLW analysis, more statistics are needed to
constraint $\gamma$ from the ADS method.

\subsection{The \KS \pip \pim Dalitz (GGSZ) Analysis:}
\begin{figure*}
\includegraphics[width=16cm]{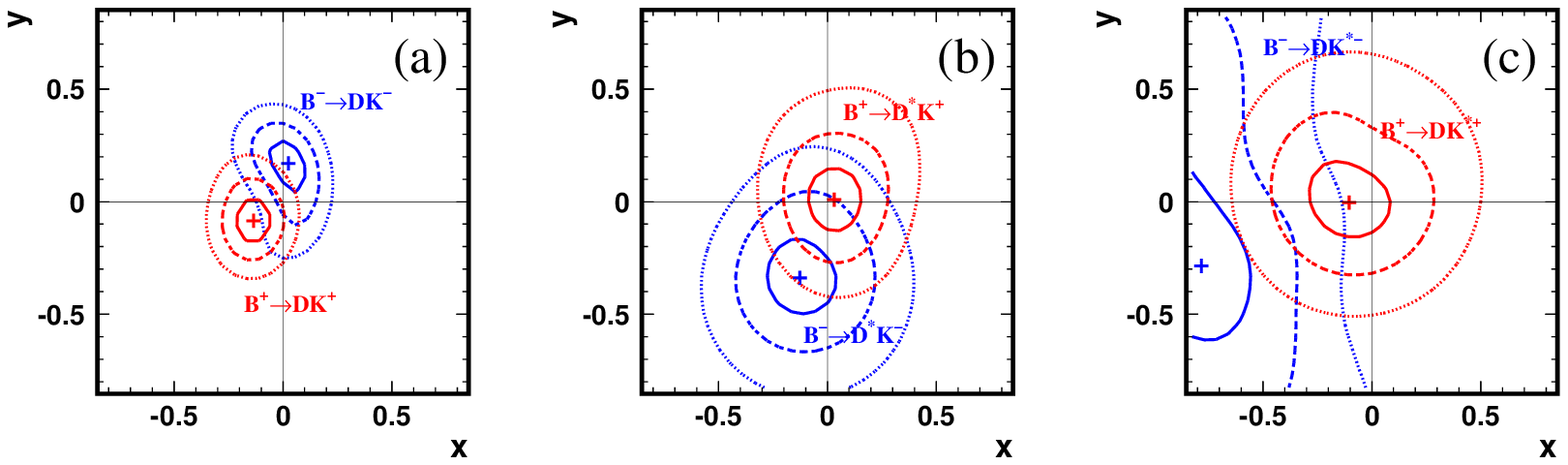}
\includegraphics[width=5.5cm]{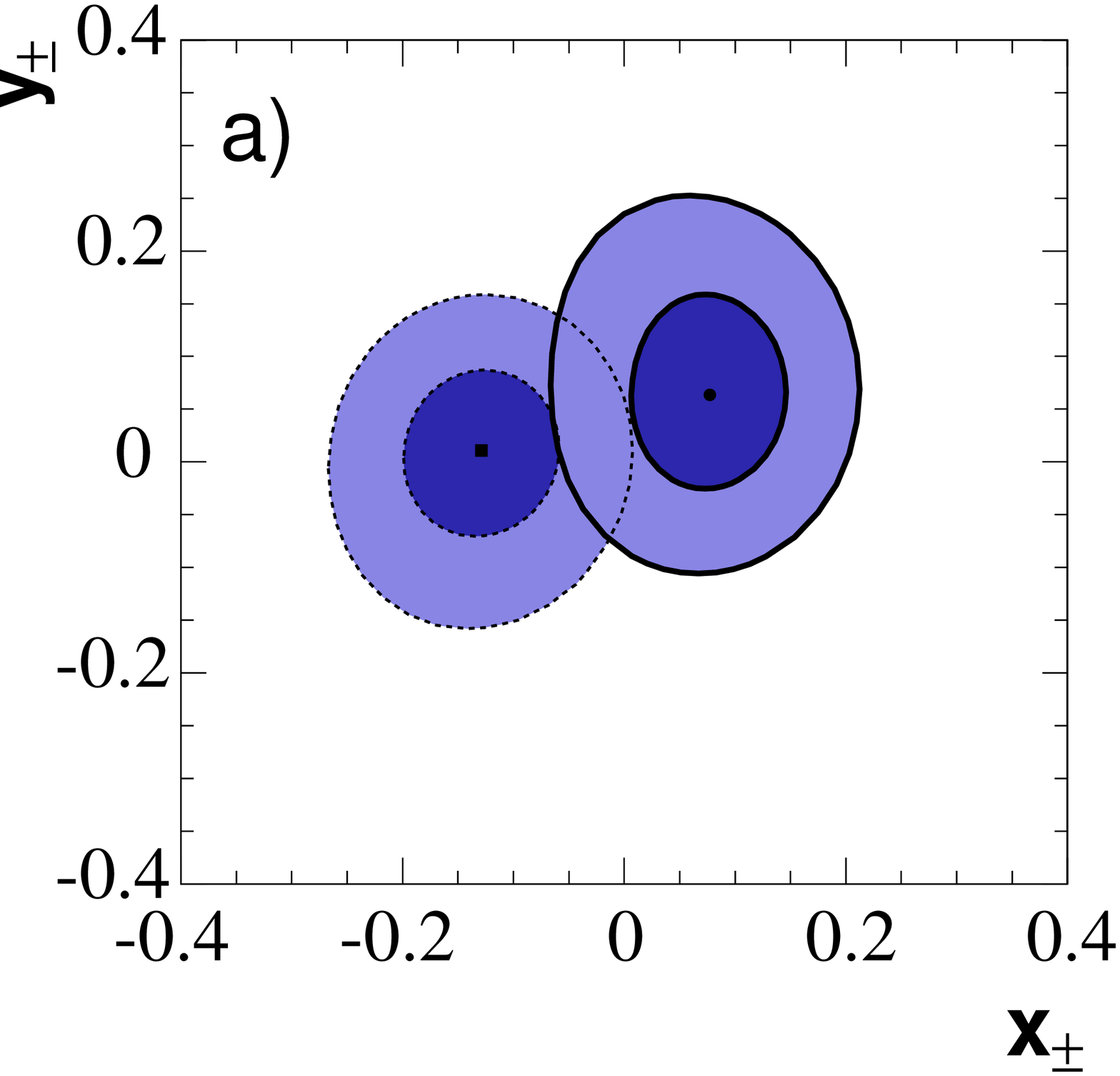}
\includegraphics[width=5.5cm]{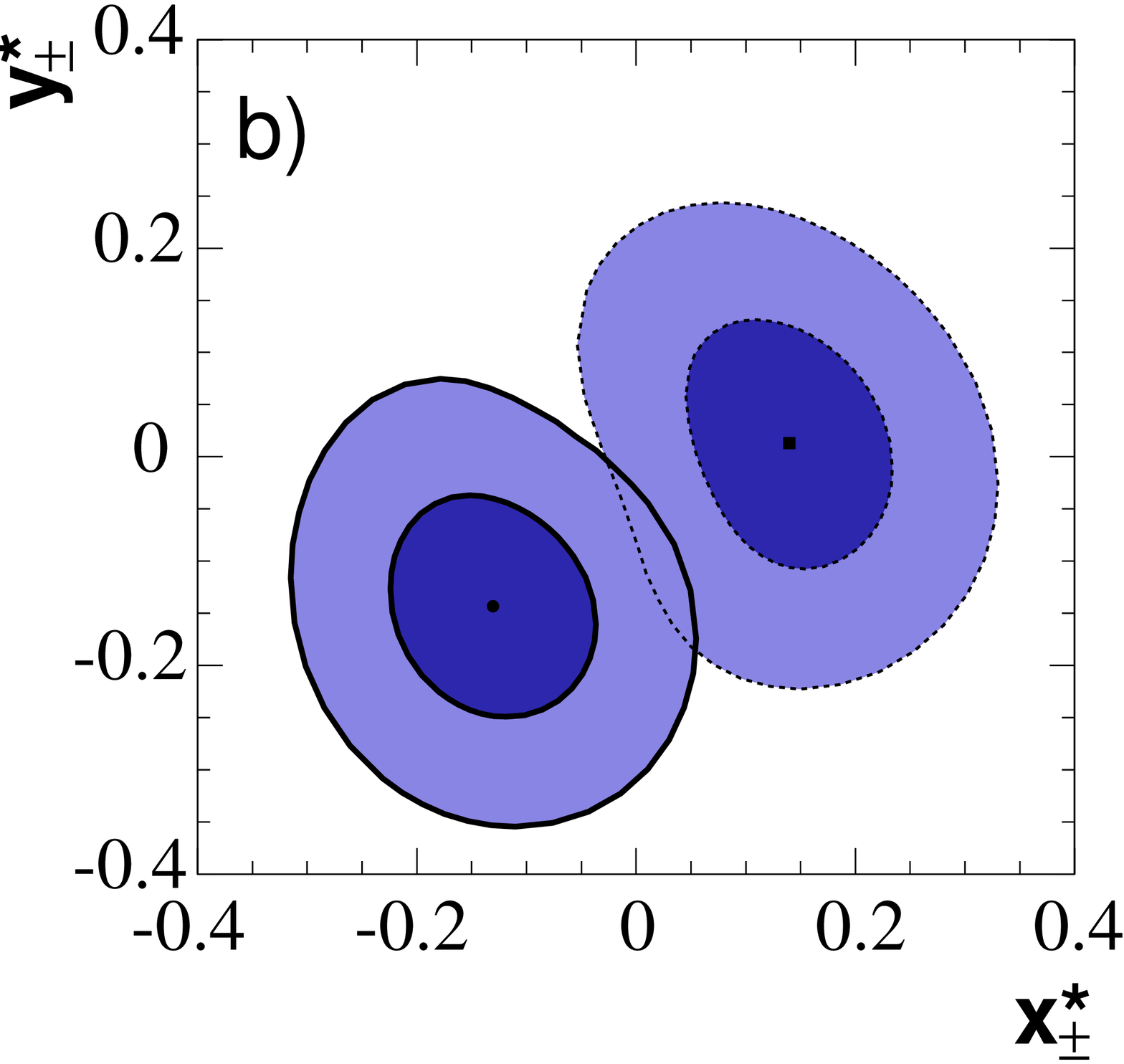}
\includegraphics[width=5.5cm]{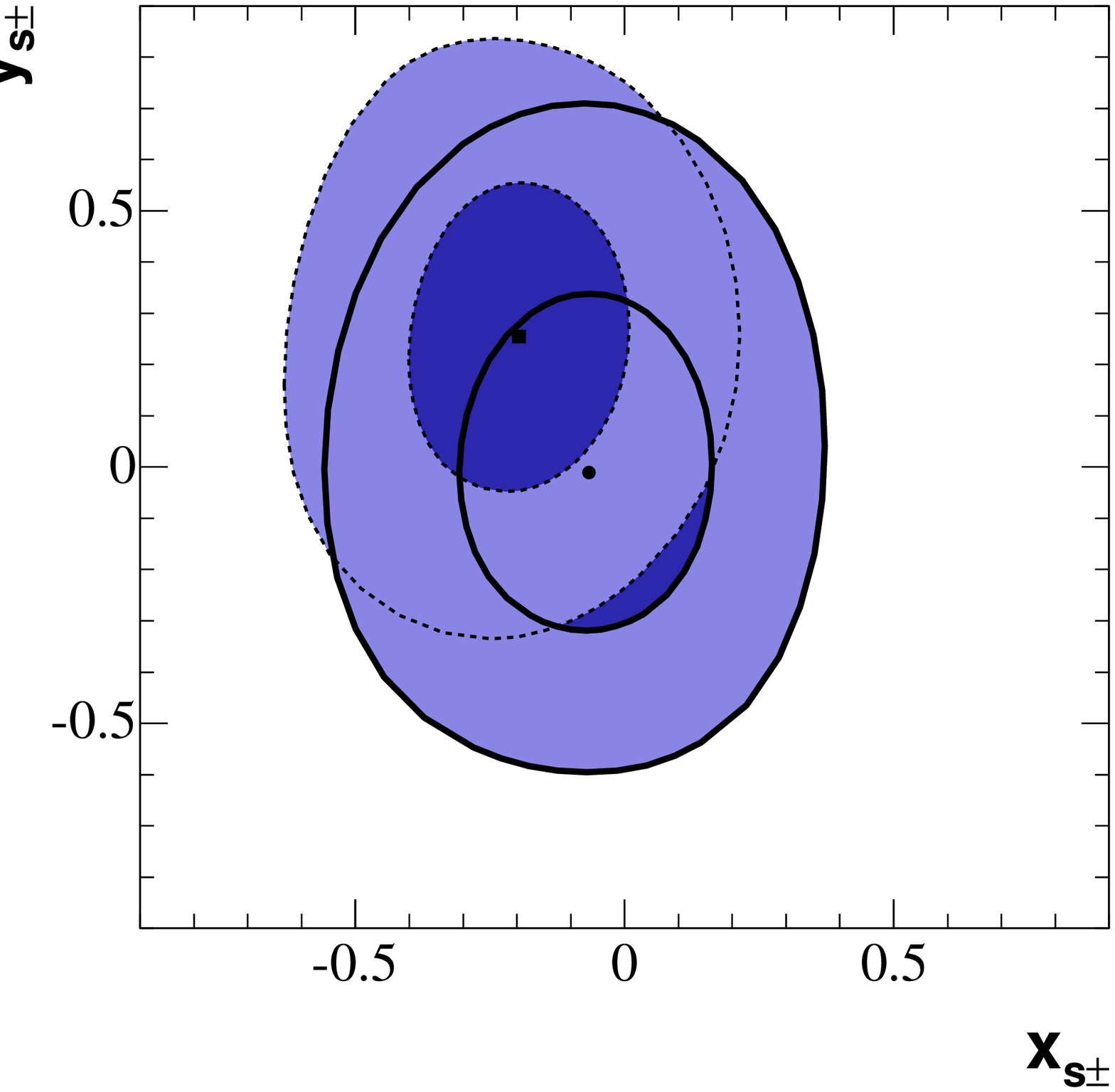}
\caption{Results of signal fits in the ($x,y$) plane for (a)
\bdkp, (b) \bdskp\ and (c) \bdksp\ samples, separately for $B^-$
           and $B^+$ data. Top: Belle 1, 2 and 3$\sigma$
           confidence contours. Blue is \Bm, red is \Bp. Bottom: \babar\ $1\sigma$(dark) and $2\sigma$ (light)
confidence contours. Solid is \Bm, dashed is \Bp.}
\label{fig:xycontours}
\end{figure*}
\begin{table*}[tb]
  \centering
  \caption{Summary of BaBar and Belle measurements of the Dalitz analysis (GGSZ)  observables $x_\pm$ and
  $y_\pm$.}
\begin{tabular}{|c|c|c|l|l|l|l|}
  \hline
  Mode & Experiment         & $N(B\overline B)$& $x_-$ $( 10^{-2})$& $y_-$ $( 10^{-2})$& $x_+$ $(10^{-2})$& $y_+$ $(10^{-2})$\\
  \hline
  $B\to D^0K$ & \babar \cite{pap:babardalitz1}    & 227$\times 10^6$ & $+8 \pm 7 \pm 3 \pm 2 $  & $+6 \pm 9 \pm 4 \pm 4 $ & $-13\pm 7 \pm 3 \pm 3$  & $+2\pm 8 \pm 2 \pm 2$ \\
              & Belle  \cite{pap:belledalitz}     & 386$\times 10^6$ & $+3^{+7}_{-8}\pm 1$      & $+17^{+9}_{- 12}\pm 2$  & $-14\pm 7 \pm 2$        & $-9\pm 9 \pm 1$  \\
              & HFAG  Avg. \cite{web:HFAG}        &  -               & $+5\pm 5 \pm 2$          & $+11\pm 7 \pm 2$        & $-14\pm 5 \pm 3$        & $-3\pm 6 \pm 2$  \\
  \hline
 $B\to D^{*0}K$ & \babar \cite{pap:babardalitz1} & 227$\times 10^6$ & $-13 \pm 9 \pm 3 \pm 2$ & $-14 \pm 11 \pm 2 \pm 3$  & $+14 \pm 9 \pm 3 \pm 3$  & $+1 \pm 12 \pm 4 \pm 6$\\
                & Belle  \cite{pap:belledalitz}  & 386$\times 10^6$ & $-13^{+17}_{- 15} \pm 2 $ & $-34 ^{+17}_{- 16}\pm 3 $  & $+3 \pm 12 \pm 1$  & $+1\pm 14 \pm 1$  \\
                & HFAG  Avg.  \cite{web:HFAG}    &  -               & $-13\pm 8 \pm 2$          & $-20 \pm 9 \pm 3$  & $+10 \pm 7 \pm 3$  & $+1 \pm 9 \pm 6$  \\
  \hline
  $B\to D^0K^*$ & \babar \cite{pap:babardalitz2} & 227$\times 10^6$ & $-20\pm 20\pm 11 \pm 3$ & $+26 \pm 30 \pm 16 \pm 3$ & $-7\pm 23\pm 13\pm 3$    & $-1 \pm 32 \pm 18 \pm 5$  \\
                & Belle  \cite{pap:belledalitz}  & 386$\times 10^6$ & $-78^{+25}_{-30}\pm 3$  & $-28^{+44}_{-34} \pm 5$   & $-11^{+18}_{-17} \pm 1$  & $0 \pm 16 \pm 1$    \\
                & HFAG  Avg.  \cite{web:HFAG}    & -   & $ -46 \pm  17 \pm  3$ & $ +5 \pm  27 \pm  3$ & $ -10 \pm  15 \pm  3$ & $ 0 \pm  15 \pm  5$   \\
  \hline
\end{tabular}
\label{tab:ggsz}
\end{table*}
A third method to constraint $\gamma$ from \Bm \to \Dz \Km decays
is to use three body final states common to \Dz and \Dzb, as
suggested in Ref.\cite{ref:GGSZ}. Among the possible \Dz 3-body
decay modes the \KS\pip\pim channel appears to have the largest
sensitivity to $\gamma$, because of the best overall combination
of branching ratio magnitude, \Dz-\Dzb interference and background
level. Defining the Dalitz plot amplitude of the decay \Dz \to
\KS\pip\pim as $A_D(m^2_-,m^2_+)$, where $m^2_-=m^2(\KS\pim)$ and
$m^2_+=m^2(\KS\pip)$, the Dalitz plot density of the \Dz in \Bpm
\to \Dz \Kpm decays can be written as

\begin{eqnarray}
& d\sigma(m^2_\mp,m^2_\pm) \equiv  |A_D(m^2_\mp,m^2_\pm)|^2 +
r_B^2
|A_D(m^2_\pm,m^2_\mp)|^2 \nonumber \\
& +  2 r_B
Re[A_D(m^2_\mp,m^2_\pm)A_D^*(m^2_\pm,m^2_\mp)e^{i(\delta_B \mp
\gamma)} ]\label{eq-dalitz}.
\end{eqnarray}
Information on the weak phase $\gamma$, the strong phase
difference $\delta_B$ and the ratio $r_B$ of the two \Bpm \to \Dz
\Kpm decay  amplitudes can be obtained from simultaneous fit of
the Daliz plot density of \Bm and \Bp data if the Dalitz decay
amplitude $A_D(m^2_\mp,m^2_\pm)$ is precisely known. Both \babar\
and Belle extract $A_D(m^2_\mp,m^2_\pm)$ by fitting high purity
tagged \Dstarp\to \Dz \pip, \Dz \to \KS\pip\pim samples using an
Isobar Model [Coherent sum of 16 (15) Breit-Wigner amplitudes for
\babar [(Belle), plus a non-resonant component] to model the \Dz
decay amplitude. The Dalitz regions corresponding to the
Doubly-Cabibbo suppressed decays \Dz\to\Kstarp(892)\pim and
\Dz\to\Kstarp(1430)\pim ["ADS like"] and to the CP-odd decays
\Dz\to\KS$\rho^0$ decays ["GLW" like] are found to have the
largest sensitivity to $\gamma$. Both \babar\ and Belle introduce
the cartesian coordinates $x_\pm = Re(r_B e^{i(\delta_B\pm
\gamma)})$ and $y_\pm = Im(r_B e^{i(\delta_B\pm \gamma)})$ to fit
their data. A total of 12 parameters ($3\times 4$) is extracted by
each experiment from the Dalitz plot density fit of the \Dz in
\Bpm\to\Dz\Kpm, \Bpm\to\Dstarz\Kpm and \Bpm\to\Dz\Kstarpm data.
\begin{figure*}[!t]
\begin{center}
 \includegraphics[width=4.9cm]{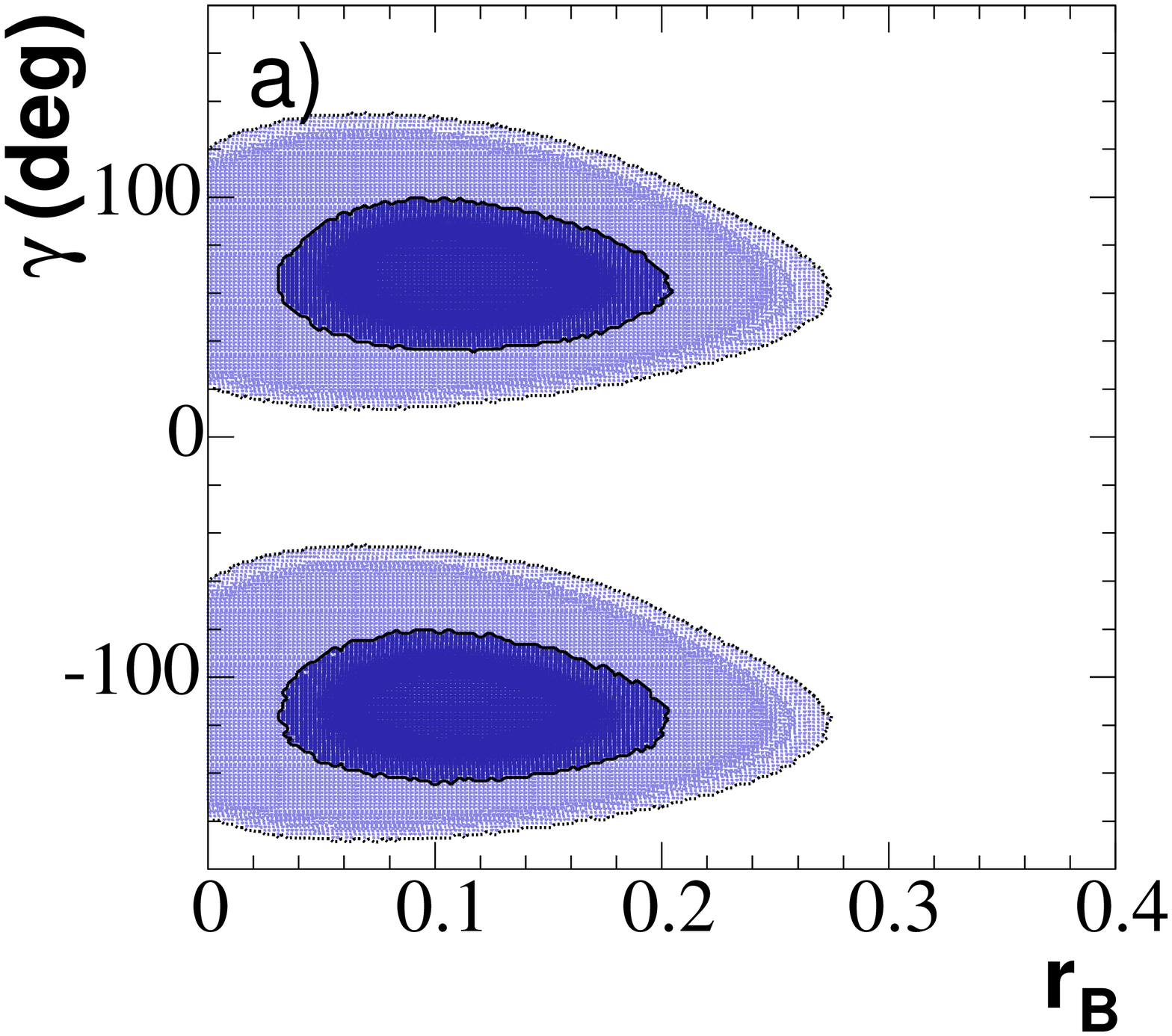}
 \includegraphics[width=4.9cm]{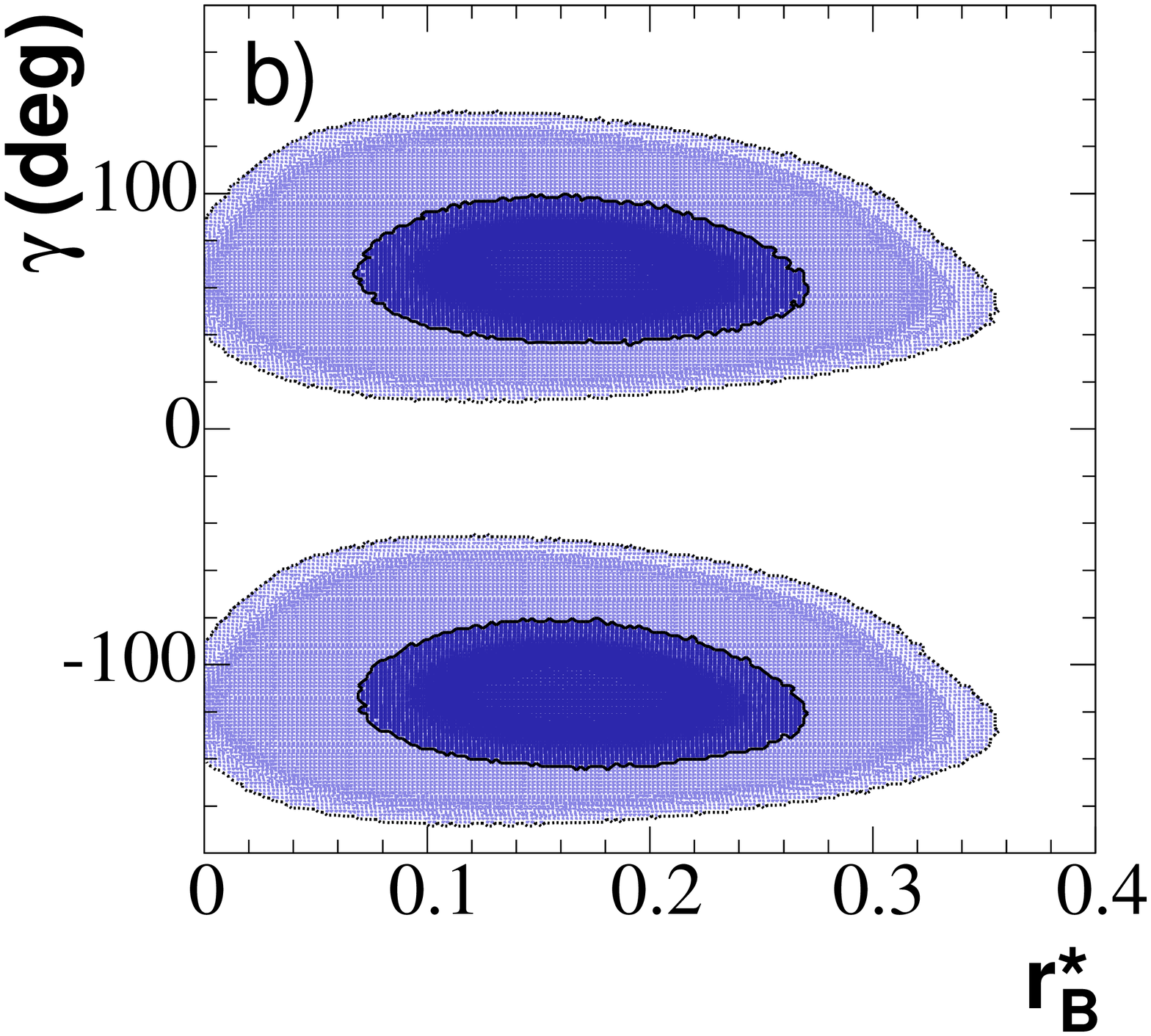}
 \includegraphics[width=4.9cm]{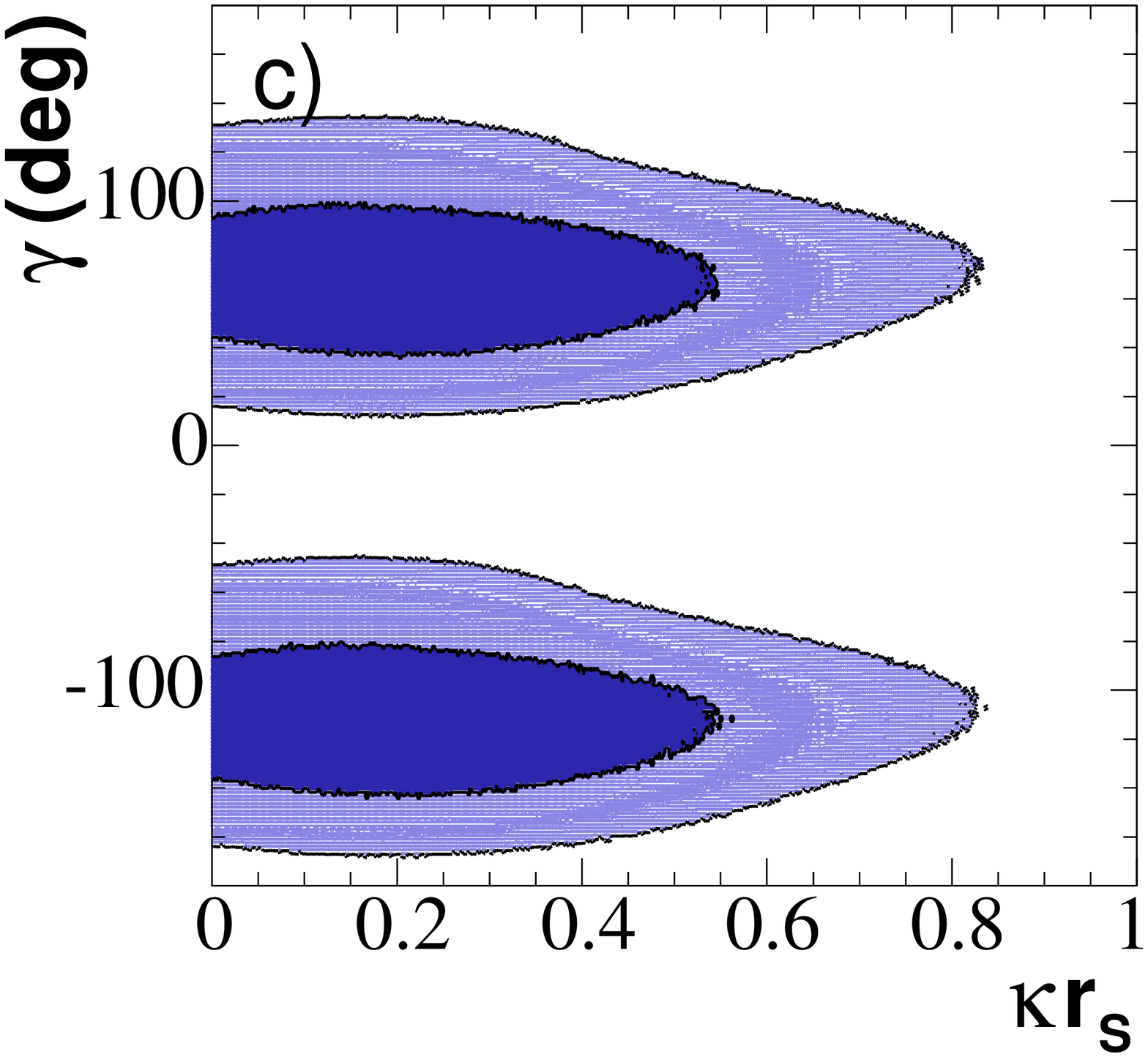}
\caption{\babar\ analysis: two-dimensional projections onto the
(a) $\rb-\g$, (b) $\rbs-\g$, and (c) $\kappa\rs-\g$ planes of the
seven-dimensional one- (dark) and two- (light) standard deviation
regions, for the combination of $\Bm \to \dodstartilde \Km$ and
$\Bm \to \Dztilde \Kstarm$ modes. } \label{fig:contours-rbvsgamma}
\end{center}
\end{figure*}

\begin{figure}[!t]
\begin{center}
\includegraphics[width=6.9cm]{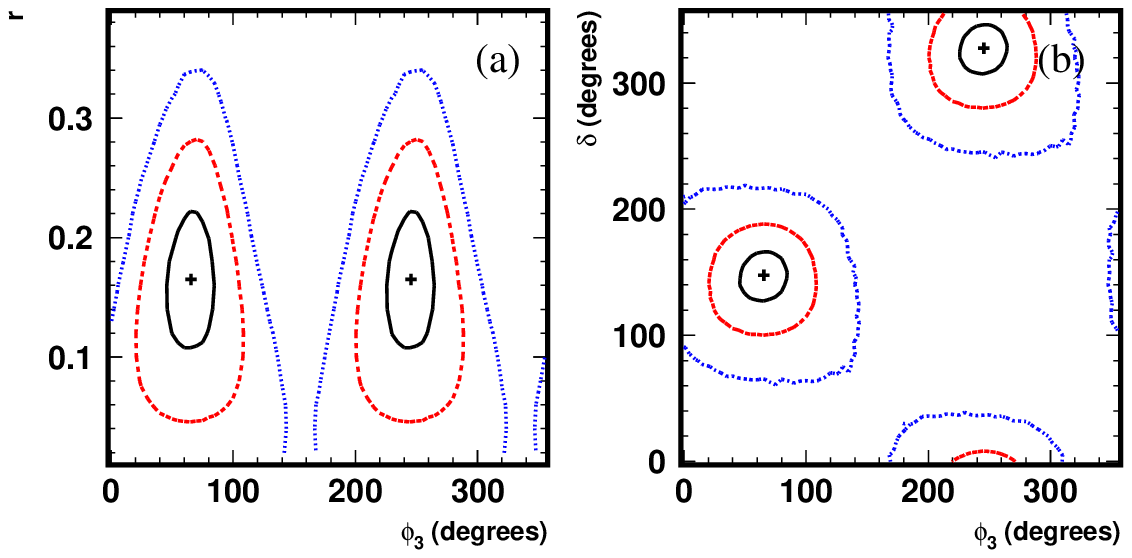}
\includegraphics[width=6.9cm]{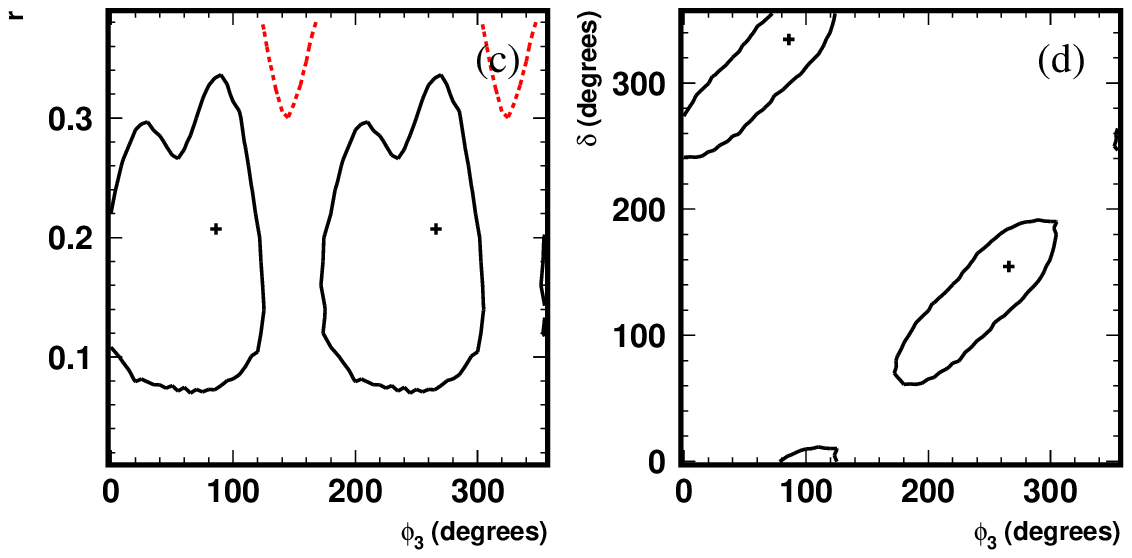}
\includegraphics[width=6.9cm]{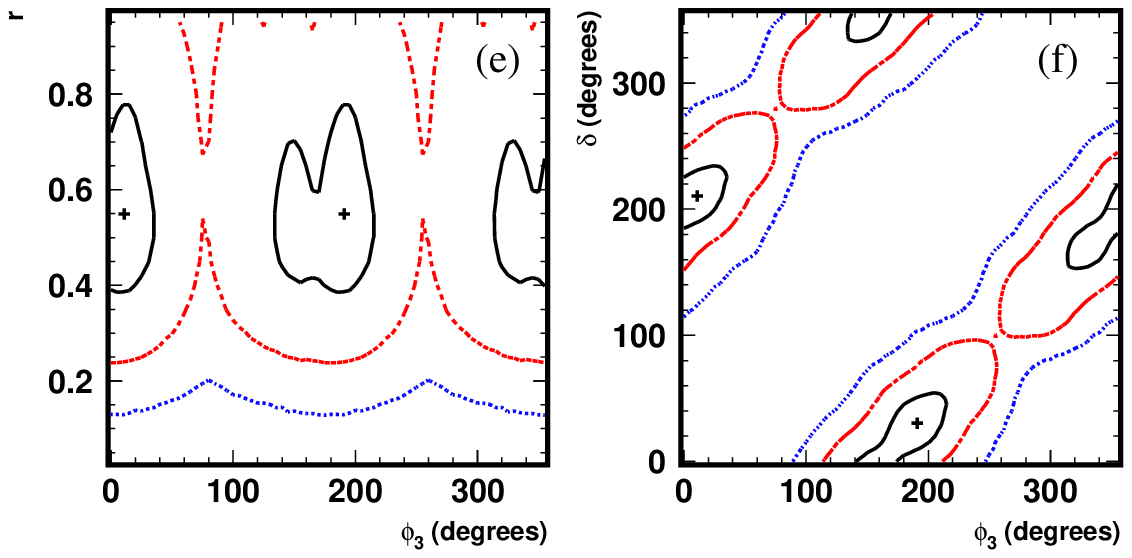}
\caption{Belle\ analysis: two-dimensional projections onto the
($\rb-\g/\phi_3$) plane (left) and the ($\delta_B-\g/\phi_3$)
plane (right) for the (a,b) \Bm\to\Dz\Km , (c,d) \Bm\to\Dstarz\Km,
and (e,f) \Bm\to\Dz\Kstarm modes. Contours indicate 1, 2 and
3$\sigma$ confidence regions.} \label{fig:bellecontours-rbvsgamma}
\end{center}
\end{figure}

\par Using a sample of 227$\times 10^6$ \BB events, \babar\
reconstructs $282\pm 20$ events in the $B\to\Dz K$ channel, $90\pm
11$ ($44\pm 8$) in the $B\to\Dstarz K$, $\Dstarz \to \Dz\piz$
($\Dstarz \to \Dz\g$) channel and $42\pm 8$ events in the
$B\to\Dz\Kstar$ channel. Belle reconstructs $331\pm 17$ $\Dz K$,
$81\pm 8$ $\Dstarz_{[\Dz\piz]}K$ and $54\pm 8$ \Dz\Kstar events
out of 357$fb^{-1}$ of data ($\sim 386\times 10^6 $ \BB events).
The results of the fits of the ($x_\pm$, $y_\pm$) parameters from
\babar\ and Belle for the different $B$ decay channels are shown
in Fig.\ref{fig:xycontours} and summarized in
Table~\ref{tab:ggsz}. A significant separation between the \Bp and
the \Bm data, indicative of direct CP violation, is visible in the
$\Dz K$ and $\Dstarz K$ data.  Both experiments use a frequentist
analysis to interpret the constraints on ($x_\pm$, $y_\pm$) in
terms of the physical parameters ($r_B$, $\delta_B$ and $\gamma$).
From the combination of the different $B$ decay channels, the
two-dimensional constraints obtained by \babar\ in the ($r_B$, \g)
plane for $DK$, $D^*K$ and $DK^*$ are shown in
Fig.\ref{fig:contours-rbvsgamma}. The value of $\gamma$ is
constrained by \babar\ to be $67^o\pm 28^o\pm 13^o\pm 11^o$, where
the first error is statistical, the second one is the experimental
systematic uncertainty and third reflects the Dalitz model
uncertainty. The values found by \babar\ for $r_B$ are
$r_B=0.12\pm 0.08 \pm 0.03\pm 0.04$ for the \bdkm\ mode,
$r_B=0.17\pm 0.10 \pm 0.03\pm 0.03$ for the \bdskm\ mode and
$r_B<0.50 (0.75)$ at one (two) standard deviation level
 for the \bdksm\ mode.
The equivalent constraints from Belle on the ($r_B$, \g) and
($\delta_B$, \g) planes are shown in
Fig.\ref{fig:bellecontours-rbvsgamma}. Belle reports
$\gamma/\phi_3=53^o\;^{+15^0}_{-18^0}\pm 3^o\pm 9^o$ and finds for
the ratio $r_B$ of the two interfering amplitudes
$r_B=0.159^{+0.054}_{-0.050}\pm 0.012\pm 0.049$ for the \bdkm\
mode, $r_B=0.175^{+0.108}_{-0.099}\pm 0.013\pm 0.049$ for the
\bdskm\ mode and $r_B=0.564^{+0.216}_{-0.155}\pm 0.041\pm 0.084$
for the \bdksm\ mode. More details on the analysis are given in
Ref.\cite{pap:babardalitz1, pap:babardalitz2,pap:belledalitz}.

\section{sin($2\beta+\gamma$) measurements}
\begin{figure}[hb]
\centering
\includegraphics[width=40mm]{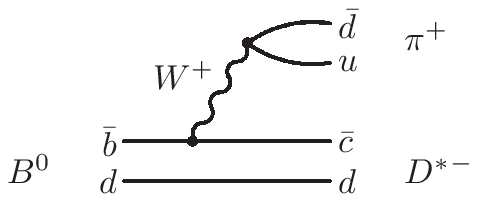}
\includegraphics[width=40mm]{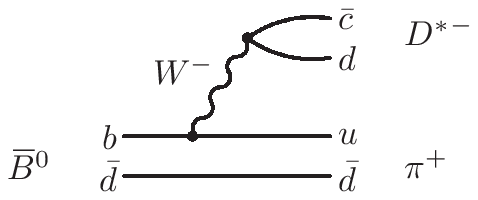}
  \caption{Feynman diagrams for the Cabibbo-favored decay $B^0 \to D^{*-} \pi^+$ (left)
  and the Cabibbo-suppressed decay $\Bzb\to D^{*-} \pi^+$ (right).
   The CKM-suppression factor is
   $|V_{cd}V_{ub}^*/V_{cb}V_{ud}^*| \propto \lambda^2 \sqrt{\rho^2+\eta^2} \approx 0.02$.}
\label{fig:feynman}
 \end{figure}

Time-dependent asymmetries in $B^0\to D^{(*)}\pi$, $D^{(*)}\rho$
and $B^0 \to D^{(*)0}K^0$ can be used to constrain
$\sin(2\beta+\gamma)$\cite{pap:dunietz2bg}. As $\beta$ is well
known from $b\to c \bar c s$, a constraint on the angle $\gamma$
follows. The $\Bz \to D^{(*)}\pi$ method uses an interference
between the usual Cabibbo-favored $b\to c$ channel and the
doubly-Cabibbo suppressed $b\to u$ channel
(Fig.\ref{fig:feynman}). These two amplitudes have a relative weak
phase of $\gamma$, and a weak phase of $2\beta$ is provided by the
\BzBzb mixing. These modes have the advantage of a "large" ($\sim
0.5\%$) branching fraction but the price to pay is the small ratio
$r$ of the suppressed to favored amplitudes,
$$ r = \left | \frac {A(\Bz \rightarrow D^{(*)+}h^-} {A(\Bzb \rightarrow D^{(*)+}h^-}\right | \propto \lambda^2 (\sim
0.02).$$ This results in very small CP-asymmetries. Moreover, the
ratio $r$ cannot be measured directly, but has to be estimated
from the measurement of ${\cal B}(\Bz \to D_s^+\pim)$, assuming
SU(3) flavor symmetry.
\par The experimental observables are the coefficients $S^\pm$ and $C$ of the
$\sin(\Delta m \Delta t)$ and $\cos(\Delta m \Delta t)$ terms in
the time dependent asymmetries of $B^0 (\bar B^0)\to D^{(*)\pm}
\pi^\mp$ (or $D^{(*)\pm}\rho^\mp$). For small values of $r$, the
parameter $S^\pm$ is given by $S^\pm \simeq 2 r \sin (2\beta
+\gamma \pm \delta)$, where $\delta$ is the strong phase
difference between the $b\to u$ and $b\to c$ decay amplitudes.
\par Potential competing CP violating effects can arise from $b\to
u$ transitions on the tag side if a Kaon is used to tag the flavor
of the other \Bz in  the event, resulting in an additional sin
term $S'^\pm=2r'\sin (2\beta +\gamma \pm \delta')$. Here, $r'$
($\delta'$) are the effective amplitude (phase) used to
parameterize the tag side interference. To account for this term,
\babar\ chooses to rewrite $S^\pm$ as $S^\pm = (a\pm c)+b$, where
$a = 2r \sin(2\beta+\gamma)\cos(\delta)$,  $c =
\cos(2\beta+\gamma)[2r \sin(\delta)+2 r' \sin(\delta')]$ and
$b=2r' \sin(2\beta+\gamma)\cos(\delta')$. This parametrization has
the advantage that the $a$ parameter does not depend on the
tagging category. On the other hand, the $c$ parameter can only be
estimated with lepton-tagged events, for which one has
$c=c^{lept}=\cos(2\beta+\gamma)[2r \sin(\delta)]$. The $b$
parameter characterizes CP violation on the tag side and does not
contribute to the interpretation.  In the approach chosen by
\babar, the $a$ and $c^{lept}$ parameters are fitted. Belle, on
the contrary, chooses to fit the $S^\pm$ parameters but measures
tag-side CPV parameters $S'^\pm$ using a sample of $\Dstar l
\nu_l$ events, which can have only tag-side CP-violation.

\begin{figure}[hb]
\centering
\includegraphics[width=5.9cm]{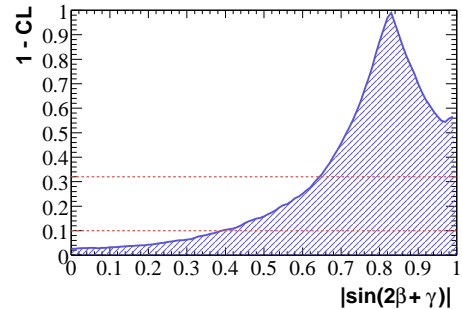}
\caption{Frequentist confidence level as a function of
$|sin(2\beta\!+\!\gamma)|$, obtained when combining the \babar\
results on exclusive decays with the result obtained on partially
reconstructed $B {\to} D^{*\pm}\pi^{\mp}$
decays~\cite{pap:babar2bg1}. The horizontal lines show the $68\%$
(top) and $90\%$ C.L. (bottom).} \label{fig:s2bg1}
\end{figure}

\par \babar\ and Belle use two experimental methods for reconstructing $\Bz
(\Bzb)\to D^{(*)}\pi$ and $D^{(*)}\rho$ decays. They perform
either exclusive reconstruction, where the hadronic decay modes
with  $\Dz \to \Km \pip$, $\Km\pip\piz$ and $\Km\pip\pim\pip$ are
fully reconstructed, or partial reconstruction of
$D^{*\pm}\pi^\mp$, where only the slow $\pi$, and not the \Dz from
$D^*\to \Dz \pi$, is reconstructed. Using only the slow pion
provides sufficient kinematic constraints to reconstruct this
decay. The fully exclusive method has a very high signal purity
[typically larger than 90\%] but a lower efficiency. The
semi-inclusive method has efficiencies 5 times larger but the
purity is only $\sim 30\%$ [$\sim 50\%$] for Kaon [Lepton] tags.
\par \babar\ has published results based on a statistics of
$232\times 10^6$ \BB events \cite{pap:babar2bg1}. From 18710
$D^{*\pm}\pi^\mp$ events tagged with a lepton (purity 54\%), and
70580 $D^{*\pm}\pi^\mp$ events tagged with a kaon (purity 31\%)
the parameters related to $2\beta+\gamma$ are measured to be
\begin{eqnarray}
a_\dstpi &=& 2r\sin(2\beta+\gamma)\cos\delta  \nonumber \\
&=& -0.034 \pm 0.014 \pm 0.009 \nonumber
\end{eqnarray}
and
\begin{eqnarray}
c_\dstpi^{lept}   &=& 2 r \cos(2\beta+\gamma)\sin\delta \nonumber \\
   &=&  -0.019 \pm 0.022 \pm 0.013,\nonumber
\end{eqnarray}
where the first error is statistical and the second is systematic.
This is the world most precise measurement of CP-violating
parameters in $B\to D^{(*)}\pi$ decays to date. \babar\ has also
published results based on a statistics of $232\times 10^6$ \BB
events for the fully exclusive analysis of $B^0\to D\pi$, $D^*\pi$
and $D\rho$ \cite{pap:babar2bg2}. From a time-dependent maximum
likelihood fit to a sample of 15038 $D^{\pm}\pi^\mp$ events
(purity 87\%), 14002 $D^{*\pm}\pi^\mp$ events (purity 87\%), and
8736 $D^{\pm}\rho^\mp$ events (purity 82\%), the parameters
related to the \CP violation angle $2\beta+\gamma$ are measured to
be:
\begin{eqnarray}
a_{D\pi}           &=& -0.010 \pm 0.023 \pm 0.007,    \nonumber \\
c_{D\pi}^{\rm lept}&=& -0.033 \pm 0.042 \pm 0.012,    \nonumber \\
a_{D^*\pi}         &=& -0.040 \pm 0.023 \pm 0.010,    \nonumber \\
c_{D^*\pi}^{\rm lept} &=& +0.049 \pm 0.042 \pm 0.015, \nonumber \\
a_{D\rho}             &=& -0.024 \pm 0.031 \pm 0.009, \nonumber \\
c^{\rm lept}_{D\rho}  &=& -0.098 \pm 0.055 \pm 0.018, \nonumber
\label{eq-babarexcl}
\end{eqnarray}
where the first error is statistical and the second is systematic.
These results are combined with the result obtained on the
partially reconstructed $B {\to} D^{*\pm}\pi^{\mp}$ sample, using
a frequentist method described in Ref.~\cite{pap:babar2bg1} to set
a constraint on $2\beta\!+\!\gamma$.
Based on the results from
Refs.\cite{pap:babardspiold,pap:babardsrho}, the values of the
amplitude ratios $r_{D^*\pi}$, $r_{D\pi}$ and $r_{D\rho}$ used to
set this constraint are:
\begin{eqnarray}
  r_{D^* \pi} & = & 0.015^{+0.004}_{-0.006}\pm 0.005 (\mathrm{theory}),
  \nonumber \\
  r_{D \pi} & = & 0.019 \pm 0.004\pm 0.006 (\mathrm{theory}),
\nonumber \\
  r_{D \rho} & = & 0.003 \pm 0.006\pm 0.001(\mathrm{theory})\nonumber
\end{eqnarray}%
The confidence level as a function of $|sin(2\beta\!+\!\gamma)|$
is shown in Fig.~\ref{fig:s2bg1} and \babar\ sets the lower limits
$|sin(2\beta\!+\!\gamma)|\!>\!0.64\ (0.40)$ at $68\%$ $(90\%)$
C.L.

\begin{figure}[htb]
  \begin{center}
    \includegraphics[width=0.235\textwidth]{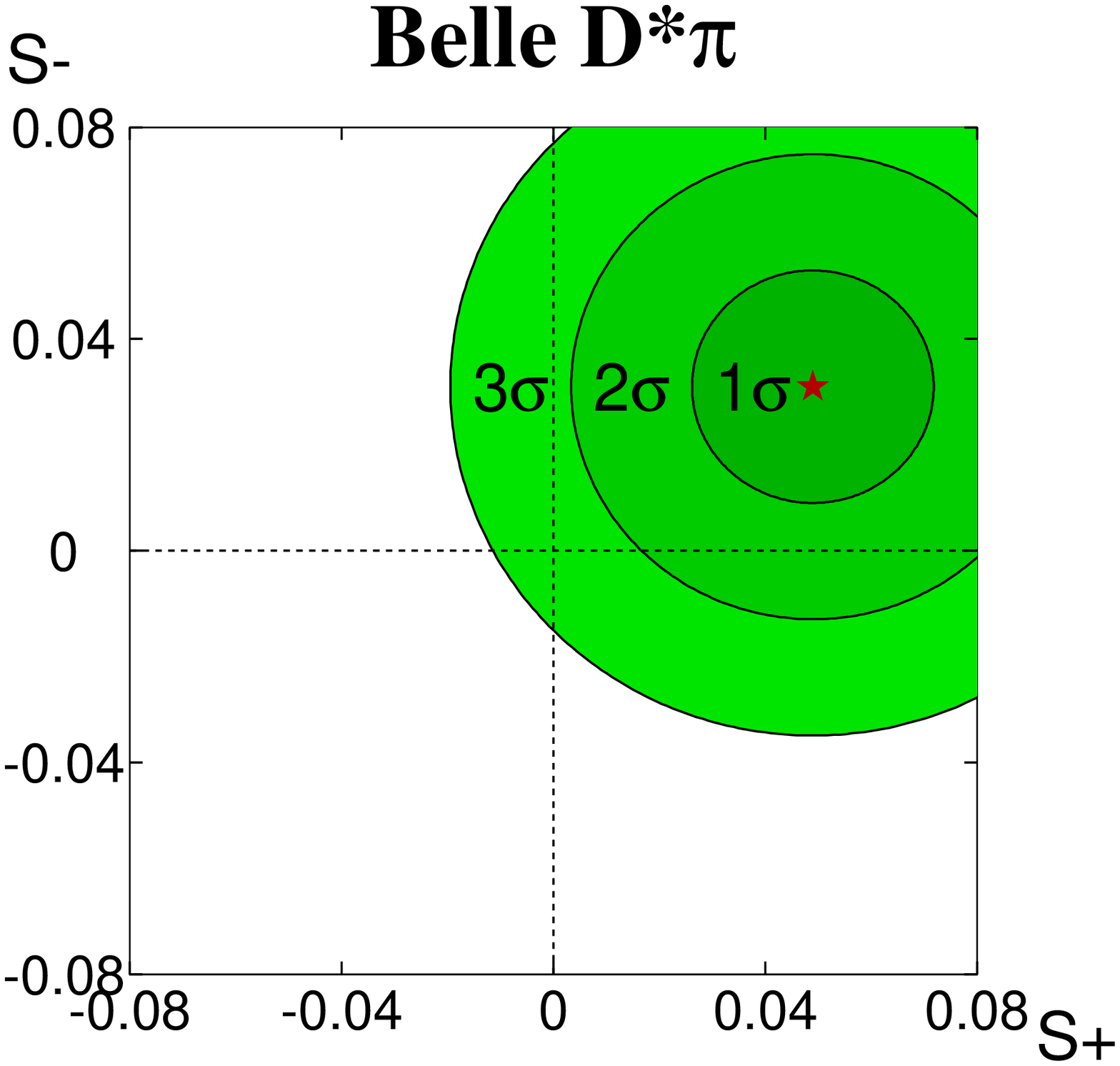}
    \includegraphics[width=0.235\textwidth]{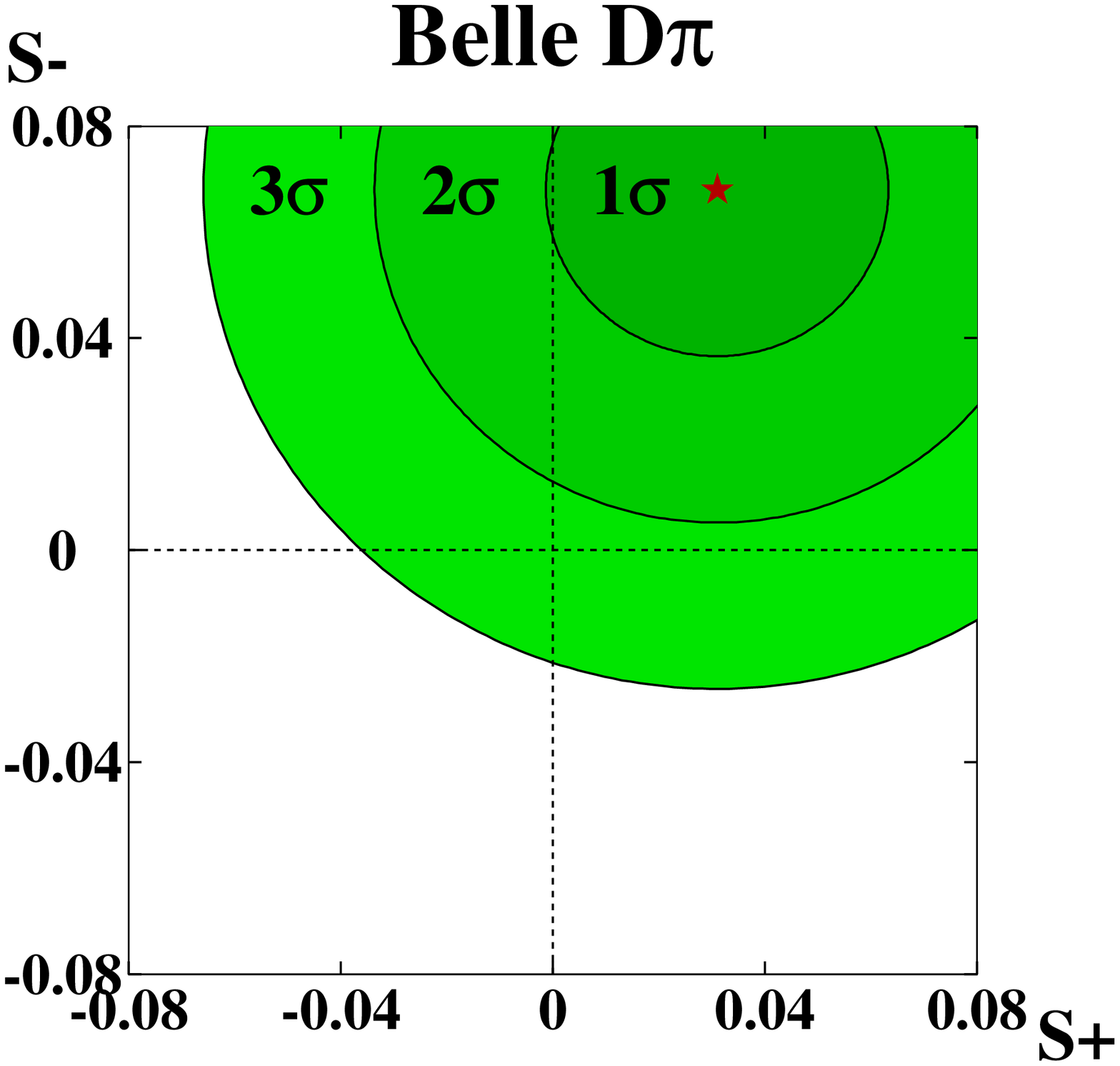}
  \end{center}
  \caption{Results of the $S^\pm$ measurements expressed in terms of
 $S^-$ vs $S^+$ for the $D^* \pi$ (left) and $D \pi$ (right) modes.
 Shaded regions indicate  allowed regions with $1\,\sigma$, $2\,\sigma$
 and $3\,\sigma$ uncertainties defined
 by $\sqrt{-2\mathrm{ln}{\cal L}} = 1,\, 4,\, 9$, respectively
}
    \label{fig:s-final}
\end{figure}

For the 2006 winter conferences, Belle has released a new result
based on an integrated luminosity of 357$fb^{-1}$, corresponding
to approximately $386\times 10^6$ \BB events \cite{pap:belle2bg}.
The CP violation parameters used in the Belle analysis are
\begin{equation}
S^{\pm} = \frac{2 (-1)^L r \sin(2\phi_1+\phi_3 \pm \delta)}
               { \left( 1 + r^2 \right)},
\label{eq:spm}
\end{equation}
where $L$ is the orbital angular momentum of the final state (1
for $D^* \pi$ and 0 for $D \pi$), and $\delta$ is the strong phase
difference of the $V_{cb}$ and $V_{ub}$ amplitudes. The values of
$r$ and $\delta$ depend on the choice of the final states, and are
denoted with subscripts $D^* \pi$ and $D \pi$ in what follows. It
should be noted that the definition of the $S^{\pm}$ parameter
used by \babar\ and Belle differ by a factor $(-1)^L$. With the
partial reconstruction method, 21741 $D^*\pi$ events tagged by a
lepton from the opposite $B$ decay are reconstructed, and the
purity is 66\%. With the fully exclusive reconstruction, 31491
$D^*\pi$ events (purity 89\%) and 31725 $D\pi$ events (purity
83\%) are reconstructed (all tags). The final results expressed in
terms of $S^+$ and $S^-$, which are related to the CKM angles
$\beta/\phi_1$ and $\gamma/\phi_3$, the ratio of suppressed to
favoured amplitudes, and the strong phase difference between them,
as $S^{\pm} = -r_{D^*\pi} \sin(2\beta+\gamma \pm \delta_{D^*\pi})/
                \left( 1 + r_{D^*\pi}^2 \right)$ for $D^* \pi$ and
$S^{\pm} = +r_{D\pi} \sin(2\phi_1+\phi_3 \pm \delta_{D\pi})/
                \left( 1 + r_{D\pi}^2 \right)$ for $D \pi$,
are
\begin{eqnarray}
S^+ (D^* \pi)&=& 0.049 \pm 0.020 \pm 0.011, \nonumber \\
S^- (D^* \pi)&=& 0.031 \pm 0.019 \pm 0.011, \nonumber \\
S^+ (D \pi)&=& 0.031 \pm 0.030 \pm 0.012, \nonumber \\
S^- (D \pi)&=& 0.068 \pm 0.029 \pm 0.012,\nonumber
\end{eqnarray}
where the first errors are statistical and the second errors are
systematic. These results are shown in Fig.~\ref{fig:s-final} in
terms of $1\,\sigma$, $2\,\sigma$ and $3\,\sigma$ allowed regions
in the $S^-$ versus $S^+$ space. They are an indication of $CP$
violation in $B^0 \to D^{*-}\pi^+$ and $B^0 \to D^- \pi^+$ decays
at the $2.5\,\sigma$ and $2.2\,\sigma$ levels, respectively.

Using published \babar\ and Belle measurements of ${\cal B}(B^0\to
D_s^{(*)+}\pi^-)$ plus some theoretical assumptions based on SU(3)
symmetry, the Belle collaboration estimates the values of the
amplitude ratios $r_{D^*\pi}$ and $r_{D\pi}$ to be
\begin{eqnarray}
  r_{D^* \pi} & = & 0.020 \pm 0.007 \pm 0.006 (\mathrm{theory}),
  \nonumber \\
  r_{D \pi} & = & 0.021 \pm 0.004 \pm 0.006 (\mathrm{theory})\nonumber
\end{eqnarray}
Using these values, they obtain 68\% (95\%) confidence level lower
limits on $|\sin (2\beta + \gamma)|$ of 0.44 (0.13) and 0.52
(0.07) from the $D^* \pi$ and $D \pi$ modes, respectively.

\begin{figure*}[htb]
  \begin{center}
    \includegraphics[width=0.45\textwidth]{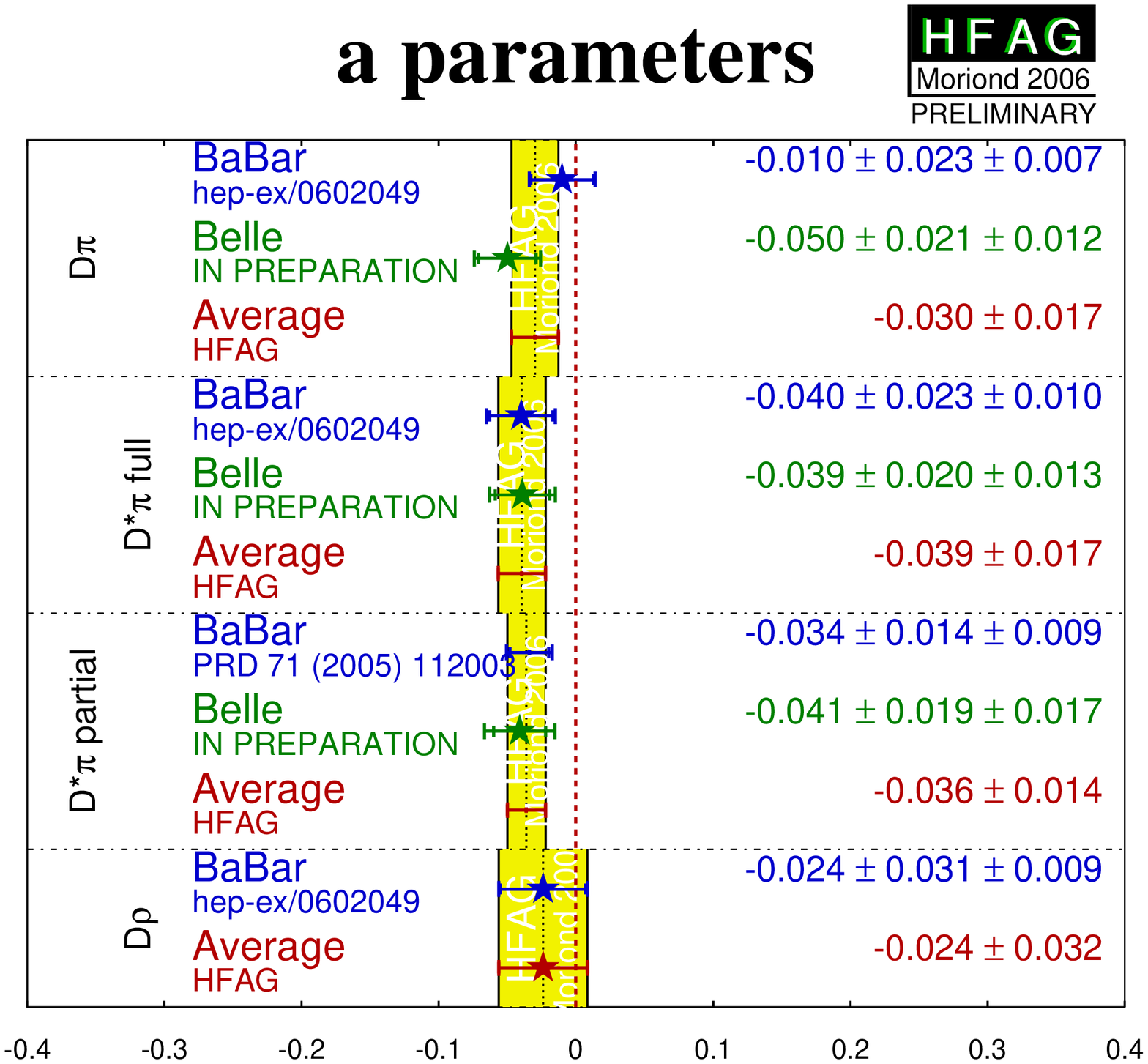}
    \includegraphics[width=0.45\textwidth]{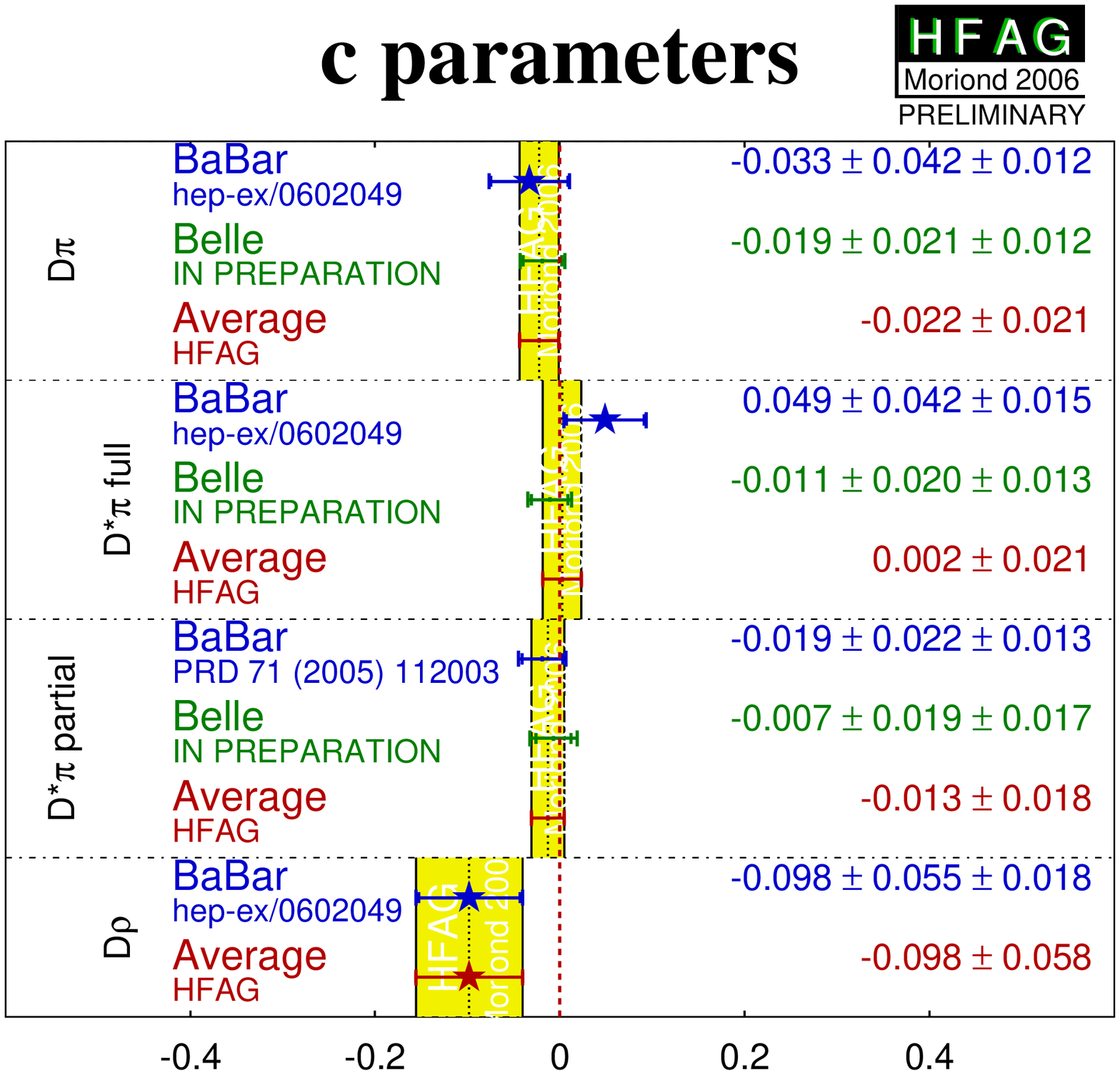}
  \end{center}
  \caption{Comparison of \babar\ and Belle sin($2\beta+\gamma$)
  results and HFAG averages for the different channels \cite{web:HFAG}.
}
    \label{fig:2bghfag}
\end{figure*}

\par In order to compare and average the \babar\ and Belle results,
the HFAG group \cite{web:HFAG} has converted the Belle results to
express them in terms of the parameters $a$ and $c$ used in the
\babar\ experiment. A comparison of these results, together with
the corresponding averages, is shown in Fig.\ref{fig:2bghfag}.
Individual measurements of $a$ with a statistical significance
better than 3$\sigma$ should be within reach before the end of the
B-factory era.

\section{${\mathbf {\cal B}(B^0\to
D_s^{(*)+}\pi^-)}$} \babar\ has recently submitted for publication
a new measurement of the $B^0\to D_s^{(*)+}\pi^-$ and $B^0\to
D_s^{(*)+}K^-$ branching fractions, based on a data sample of
$230\times 10^6$ \BB events \cite{pap:babardspi}. As explained in
the previous section, the decay $B^0\to D_s^{(*)+}\pi^-$ is
interesting because it can be used to estimate the suppressed to
favored amplitude ratios $r_{D^{(*)}\pi}$ in the $2\beta+\gamma$
analysis of the $B^0\to D^{(*)+}\pi^-$ channel
\cite{pap:dunietz2bg}:
\begin{equation}
r_{D^{(*)}\pi} =
  \tan\theta_c\,
  \frac{f_{D^{(*)}}}{f_{D^{(*)}_s}}\sqrt{\frac{\BR(\btodsospi)}{\BR(\btodospi)}}
  \ ,
\label{eq:rDPi}
\end{equation}
where $\theta_c$ is the Cabibbo angle, and
$f_{D^{(*)}}/f_{D^{(*)}_s}$ is the ratio of $D^{(*)}$ and
$D^{(*)}_s$ meson decay constants~\cite{fdsd}. Other
SU(3)-breaking effects are believed to affect $r_{D^{(*)}\pi}$ by
less than $30\%$.
\par From the number of signal events observed, \babar\ computes
the following branching fractions:
\begin{eqnarray}
 {\cal B}(B^0\to D_s^{+}\pi^-) & = & (1.3\pm 0.3\pm 0.2)\times 10^{-5} \nonumber \\
{\cal B}(B^0\to D_s^{*+}\pi^-)  & = & (2.8\pm 0.6\pm 0.5)\times 10^{-5}\nonumber \\
{\cal B}(B^0\to D_s^{+}K^-)& = & (2.5\pm 0.4\pm 0.4)\times 10^{-5}\nonumber \\
{\cal B}(B^0\to D_s^{*+}K^-) & = & (2.0\pm 0.5\pm 0.4)\times
10^{-5} \nonumber
\end{eqnarray}
Assuming SU(3) relation, Eq.~(\ref{eq:rDPi}), the following values
of the amplitude ratio $r$ are determined:
\begin{eqnarray}
 r_{D\pi} & = &
(1.3\pm0.2\stat\pm0.1\syst)\times10^{-2}\nonumber \\
r_{D^{*}\pi} & = &
(1.9\pm0.2\stat\pm0.2\syst)\times10^{-2}.\nonumber
\end{eqnarray}
This implies small $C\! P$ asymmetries in $\Bz{\to}
D^{(*)\mp}\pi^\pm$ decays.
\section{Search for ${\mathbf B^0\to
D_s^{(*)+}a_{0(2)}^-}$} 
\begin{figure}[hb]
\centering
\includegraphics[width=40mm]{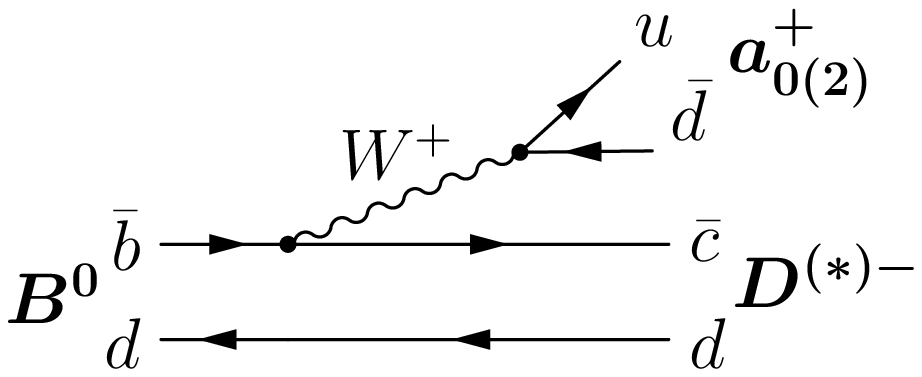}
\includegraphics[width=40mm]{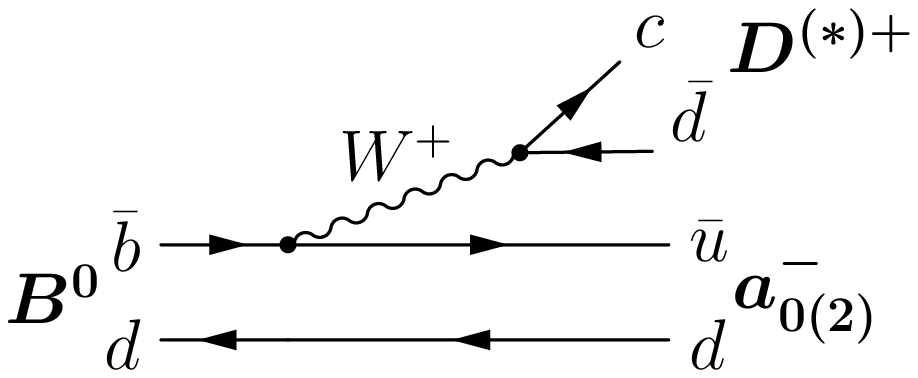}
\includegraphics[width=40mm]{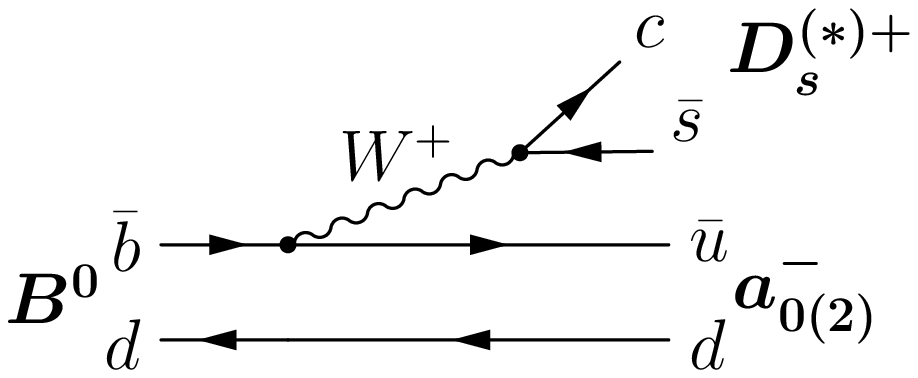}
\caption{Top diagrams: tree diagrams contributing to the decay
amplitude of $B^0 \to D^{(*)-} a^+_{0(2)}$ (including the $B^0
\Bbar^0$ mixing mediated part of the amplitude). Bottom diagram:
tree diagram representing the decay amplitude of $B^0 \to
D^{(*)+}_s a^-_{0(2)}$.} \label{fig:feynmandsa0}
 \end{figure}

It was recently suggested to use the
decays $B^0\to D^{(*)+}a_{0(2)}^-$ for measuring
sin($2\beta+\gamma$) \cite{pap:dsa0th}. These decay can proceed
through the two diagrams shown in Fig.\ref{fig:feynmandsa0} and it
is expected that the $V_{cb}$ amplitude is significantly
suppressed respective to the $V_{ub}$ amplitude, giving
significant CP-asymmetries.

\par The $V_{ub}$-mediated part of the $B^0 \to D^{(*)+} a_{0(2)}^-$
decay amplitude can be related to $B^0 \to D^{(*)+}_s a_{0(2)}^-$
using $\tan{(\theta_{\rm Cabibbo})} = |V_{cd}/V_{cs}|$ and the
ratio of the decay constants $f_{D_s^{(*)}}/f_{D^{(*)}}$.
Branching fractions of $B^0 \to D^{(*)+}_s a_2^-$ are predicted to
be in the range 1.3--1.8 (2.1--2.9) in units of
$10^{-5}$~\cite{klo}. Branching fraction estimates for $B^0 \to
D^{(*)+}_s a_0^-$ of approximately $8 \times 10^{-5}$ are obtained
using $SU(3)$ symmetry from the predictions made for $B^0 \to
D^{(*)+} a_0^-$ in~\cite{dh}.

\babar\ finds no evidence for these decays and set upper limits at
90\% C.L. on the branching fractions \cite{pap:babardsa0}: ${\cal
B}(B^0\to D_s^+ a_0^-) < 1.9\times 10^{-5}$, ${\cal B}(B^0\to
D_s^{*+} a_0^-) < 3.6\times 10^{-5}$, ${\cal B}(B^0\to D_s^+
a_2^-) < 1.9 \times 10^{-4}$, and ${\cal B}(B^0\to D_s^{*+} a_2^-)
< 2.0\times 10^{-4} $. These upper limits suggest that the
branching ratios of $B^0 \to D^{(*)+} a_{0(2)}^-$ are too small
for $CP$-asymmetry measurements given the present statistics of
the $B$-factories.

\section{Study of ${\mathbf B^0\to
D^{(*)0}K^{(*)0}}$}
\begin{figure}[htb]
\begin{center}
\includegraphics[width=0.49\linewidth]{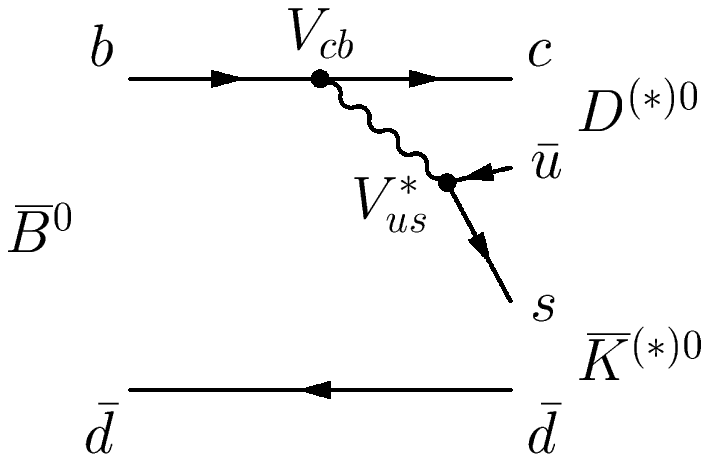}
\includegraphics[width=0.49\linewidth]{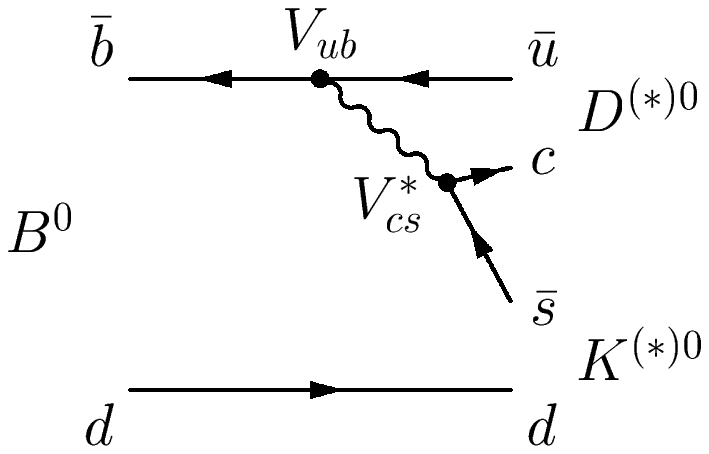}
\end{center}
\caption{ The decay diagrams for the $b\ra c$ transition
$\Bzb\ra\DDstarz\Kzb$ and the $\bbar \ra \ubar$ transition
$\Bz\ra\DDstarz\Kz$. } \label{fig:feyn}
\end{figure}

The decay modes $\Bzb\ra\DDstarz\Kzb$ offer a new approach for the
determination of \stwobg\ from the measurement of time-dependent
\CP\ asymmetries in these decays.
The \CP\ asymmetry appears as a  result of the interference
between two diagrams leading to the same final state
$\DDstarz\KS$~(Figure~\ref{fig:feyn}). A $\Bzb$ meson can either
decay via a $b\ra c$ quark transition to the $\DDstarz\Kzb$
($\Kzb\to\KS$) final state, or oscillate into a $\Bz$ which then
decays via a $\bar b\ra \bar u$ transition to the $\DDstarz\Kz$
($\Kz\to\KS$) final state.
The $\Bzb\Bz$ oscillation provides the weak phase $2\beta$ and the
relative  weak phase between the two decay diagrams is $\gamma$.

The sensitivity of this method depends on the rates for these
decays and the ratio $r_B^{(*)}$ of the decay amplitudes,
$r^{(*)}_B\equiv|{\mathcal A}(\Bzb\ra\DDstarzb\Kzb)/{\mathcal
A}(\Bzb\ra\DDstarz\Kzb)|$. In the Standard Model $r^{(*)}_B= f
\cdot  |V_{ub} V_{cs}^{*}| / |V_{cb} V_{us}^{*}|\approx f \cdot
0.4$, where the factor $f$ accounts for the difference in the
strong interaction dynamics between the  $b\ra c$ and  $b\ra u$
processes. There are no theoretical calculations or experimental
constraints on $f$. A direct determination of $r^{(*)}_B$ from the
measured rates for $\Bzb\ra\DDstarz\Kzb$ ($\Kzb\to\KS$) decays is
not possible, because one cannot distinguish between
$\Bzb\ra\DDstarz\Kzb$ and $\Bz\ra\DDstarz\Kz$. However, insight
into the $B$ decay dynamics affecting $r^{(*)}_B$ can be gained by
measuring a similar amplitude ratio $\rbk\equiv|{\mathcal
A}(\Bzb\ra\Dzb\Kstarzb)/{\mathcal A}(\Bzb\ra\Dz\Kstarzb)|$ using
the self-tagging decay $\Kstarzb\ra\Km\pip$.
\par \babar\ has recently submitted for publication a new
measurement of the $B^0\to D^{(*)0}\KS$, $\Bzb \to D^{0}\bar
K^{*0}$ and $\Bzb \to \bar D^{0}\bar K^{*0}$ branching fractions,
based on a data sample of $226\times 10^6$ \BB events
\cite{pap:babard0k0}. Defining
$\BR(\Bztilde\ra\Dstarz\Kztilde)\equiv ( \BR(\Bzb\ra\Dstarz\Kzb) +
\BR(\Bz\ra\Dstarz\Kz) )/2$ and $\BR(\Bztilde\ra\Dz\Kztilde)\equiv
( \BR(\Bzb\ra\Dz\Kzb) + \BR(\Bz\ra\Dz\Kz) )/2$, the results of
this measurement are:
\begin{eqnarray}
\BR(\Bztilde\ra\Dz\Kztilde) &=& (5.3\pm 0.7 \pm 0.3)\brscale \nonumber \\
\BR(\Bztilde\ra\Dstarz\Kztilde) &=& (3.6 \pm 1.2 \pm 0.3)\brscale \nonumber  \\
\BR(\Bzb\ra\Dz\Kstarzb) &=& (4.0 \pm 0.7
\pm 0.3)\brscale \nonumber \\
\BR(\Bzb\ra\Dzb\Kstarzb) &<& 1.1 \times 10^{-5}\ {\mathrm {at}}\
90\%\ \mathrm{C.L.} \nonumber
\end{eqnarray}

This measurement is in good agreement with previous results from
Belle \cite{pap:belled0k0}. From the absence of signal for the
$V_{ub}$ mediated mode $\Bzb\ra\Dzb\Kstarzb$, the limit $\rbk <
0.40$ at 90\% C.L.\ is obtained. The present signal yields
combined with this limit on \rbk\ suggest that a substantially
larger data sample is needed for a competitive time-dependent
measurement of \stwobg\ in $\Bztilde\ra\DDstarz\Kztilde$ decays.

\section{Conclusion}
Although the angle $\gamma/\phi_3$ is the most difficult to
measure of the Unitarity Triangle angles at the B-factories, very
promising progress has been made in constraining it over the past
few years. With the increase of statistics expected between now
and 2008, and because these measurements are theoretically clean,
both $\gamma$ and sin($2\beta+\gamma$) will progress toward
becoming precision measurements before the end of the decade.

\bigskip 
\begin{acknowledgments}
As a member of the \babar\ collaboration, I would like to thank my
\babar\ and PEP-II colleagues for their support and for choosing
me to present these results. I want also to thank Karim Trabelsi
for useful discussions about the Belle results. Finally,  I wish
to thank the organizers for a successful and very enjoyable
conference.
\end{acknowledgments}

\bigskip 

\end{document}